\documentclass{aastex61}

\submitjournal{ApJL}

\shorttitle{The largest gas filament in our Galaxy, or a new spiral arm?}
\shortauthors{Li et al.}

\begin{document}

\title{The discovery of the largest gas filament in our Galaxy, or a new spiral arm?}

\correspondingauthor{Keping Qiu}
\email{kpqiu@nju.edu.cn}

\author{Chong Li}
\affil{School of Astronomy and Space Science, Nanjing University, 163 Xianlin Avenue, Nanjing 210023, People's Republic of China}
\affil{Key Laboratory of Modern Astronomy and Astrophysics (Nanjing University), Ministry of Education, Nanjing 210023, People's Republic of China}

\author{Keping Qiu}
\affil{School of Astronomy and Space Science, Nanjing University, 163 Xianlin Avenue, Nanjing 210023, People's Republic of China}
\affil{Key Laboratory of Modern Astronomy and Astrophysics (Nanjing University), Ministry of Education, Nanjing 210023, People's Republic of China}

\author{Bo Hu}
\affil{School of Astronomy and Space Science, Nanjing University, 163 Xianlin Avenue, Nanjing 210023, People's Republic of China}
\affil{Key Laboratory of Modern Astronomy and Astrophysics (Nanjing University), Ministry of Education, Nanjing 210023, People's Republic of China}

\author{Yue Cao}
\affil{School of Astronomy and Space Science, Nanjing University, 163 Xianlin Avenue, Nanjing 210023, People's Republic of China}
\affil{Key Laboratory of Modern Astronomy and Astrophysics (Nanjing University), Ministry of Education, Nanjing 210023, People's Republic of China}


\begin{abstract}

Using the Five-hundred-meter Aperture Spherical radio Telescope (FAST), we detect a giant \ion{H}{1} filamentary structure in the sky region of 307$.\!\!^{\circ}$7 $<$ $\alpha$ $<$ 311$.\!\!^{\circ}$0 and 40$.\!\!^{\circ}$9 $<$ $\delta$ $<$ 43$.\!\!^{\circ}$4. The structure has a velocity range of $-$170 km s$^{-1}$ to $-$130 km s$^{-1}$, and a mean velocity of $-$150 km s$^{-1}$, putting it to a Galactocentric distance of 22 kpc. The \ion{H}{1} structure has a length of 1.1 kpc, which appears to be so far the furthest and largest giant filament in the Galaxy and we name it Cattail. Its mass is calculated to be 6.5 $\times$ 10$^4$ M$_{\odot}$ and the linear mass density is 60 M$_{\odot}$ pc$^{-1}$. Its width is 207 pc, corresponding to an aspect ratio of 5:1. Cattail possesses a small velocity gradient (0.02 km s$^{-1}$ pc$^{-1}$) along its major axis. Together with the HI4PI data, we find that Cattail could have an even larger length, up to 5 kpc. We also identify another new elongated structure to be the extension into the Galactic first quadrant of the Outer Scutum-Centaurus (OSC) arm, and Cattail appears to be located far behind the OSC. The question about how such a huge filament is produced at the extreme Galactic location remains open. Alternatively, Cattail might be part of a new arm beyond the OSC, though it is puzzling that the structure does not fully follow the warp of the Galactic disk.

\end{abstract}





\section{Introduction}\label{intro:}

It has been realized that filaments are one of the basic structures in the interstellar medium (ISM) \citep{2009ApJ...700.1609M,2014prpl.conf...27A}. With high spatial resolutions and sensitivities in the sub-millimeter regime, Herschel observations reveal the ubiquitous presence of filaments in the dense parts of molecular clouds, which are also most closely related to star formation \citep{2010A&A...518L.102A,2010A&A...518L.103M,2012A&A...540L..11S,2013A&A...551C...1S,2018A&A...620A..62G,2019A&A...622A..52Z}. The largest elongated molecular cloud structures are called giant (molecular) filaments with lengths greater than 10 pc. \citet{2010ApJ...719L.185J} have identified an archetype giant filament ``Nessie'' with an extreme aspect ratio (80 pc $\times$ 0.5 pc) and claimed the dense cores within it to be the birthplaces of massive stars. \citet{2014ApJ...797...53G} claimed that Nessie has a much larger length of 430 pc and referred it as the “bone” of the Scutum-Centaurus arm. After Nessie, a large number of observational studies aimed at identifying and characterizing giant filaments have been performed in various tracers, ranging from extinction maps at mid-infrared \citep{2014A&A...568A..73R,2015ApJ...815...23Z} to far-infrared/submillimeter dust emission \citep{2015MNRAS.450.4043W,2016A&A...590A.131A} and CO line emissions \citep{2015ApJ...811..134S,2017ApJ...838...49X,2018ApJS..238...10L,2020ApJS..249...27L}.

\citet{2018ApJ...864..153Z} have performed a comprehensive analysis of the physical properties of large-scale filaments in the literature. Possibly related to the mechanisms shaping the structure and dynamics of the Milky Way \citep{2014MNRAS.441.1628S}, together with the spatial resolution and sensitivity limitations of the observations, no giant filament is found in the Extreme Outer Galaxy (EOG) where the Galactocentric distance (R$_{gc}$) is greater than 15 kpc. 85\% of giant filaments lie mostly parallel to ($<$ 45$^{\circ}$), and in close proximity to ($<$ 30 pc), the Galactic plane. There are 30\%-45\% of giant filaments being associated with spiral arms. The lengths of the giant filaments range from 11 pc to 269 pc and the furthest filament is located at R$_{gc}$$\sim$12 kpc, well within the most remote spiral arms (R$_{gc}$$\sim$20 kpc) \citep{2011ApJ...734L..24D,2015ApJ...798L..27S,2016ApJ...823...77R} and the furthest molecular clouds (R$_{gc}$$\sim$30 kpc) \citep{1994ApJ...422...92D,2017PASJ...69L...3M}.


Compared to giant molecular filaments, \ion{H} {1} filaments are not well studied. Using the large-scale \ion{H} {1} surveys GASS \citep{2009ApJS..181..398M} and THOR \citep{2016A&A...595A..32B,2020A&A...634A..83W}, \citet{2016ApJ...821..117K} and \citet{2020A&A...642A.163S} systemically searched for \ion{H} {1} filaments in the Galaxy, respectively. Most of the \ion{H} {1} filaments are aligned with the Galactic plane, which is similar to the situation of giant molecular filaments. Combined with the Planck all-sky map of the linearly polarized dust emission at 353 GHz, \citet{2016ApJ...821..117K} found that \ion{H} {1} filaments tend to be associated with dust ridges and aligned with magnetic fields. They claim that the cold neutral medium (CNM) is mostly organized in sheets and they are observed as filaments due to the projection effects. The \ion{H} {1} filaments are normally cold with a typical excitation temperature T$_{ex}$$\sim$50 K and often associated with CO dark molecular gas \citep{2020A&A...639A..26K}. However, detailed  physical properties of \ion{H} {1} filaments, as well as their distribution in the Galaxy, are not well characterized. A specific case is presented in \citet{2020A&A...642A.163S}: using the 40$\arcsec$ resolution observations in the THOR survey, they identified a very long \ion{H}{1} filament ``Magdalena'' in the inner Galaxy (R$_{gc}$$\sim$12 kpc) with the length exceeding 1 kpc (see Figure 9 in their paper).

In this work we present Five-hundred-meter Aperture Spherical radio Telescope (FAST) observations of a newly detected giant filamentary \ion{H} {1} structure ``Cattail'', which is possibly the furthest (R$_{gc}$$\sim$22 kpc) and largest ($\sim$1.1 kpc) filament to date. Together with the archival HI4PI data, we find that Cattail could have an even much larger length of $\sim$5 kpc. A new extension of the Outer Scutum-Centaurus (OSC) arm between 70$^{\circ}$ $<$ $l$ $<$ 100$^{\circ}$ is also identified. The observations are described in Section \ref{Observations} and the results and discussion are presented in Section \ref{Results}.

\section{Observation and data reduction}\label{Observations}

\subsection{FAST Data}

Using the FAST, the sky region of Right Ascension of 307$.\!\!^{\circ}$7 $<$ $\alpha$ $<$ 311$.\!\!^{\circ}$0 and Declination of 40$.\!\!^{\circ}$9 $<$ $\delta$ $<$ 43$.\!\!^{\circ}$4 was observed on 2019 August 24.  The observations of \ion{H}{1} 1420.4058 MHz transition were carried out with the 19-beam L band receiver \citep{2018IMMag..19..112L}. The ROACH back-end contains  65536 channels in 32 MHz bandwidth, corresponding to a channel resolution of 0.476 kHz and a velocity resolution of 0.1 km s$^{-1}$ at 1.420 GHz. The sky region was scanned along the Right Ascension in multibeam On-the-fly (OTF) mode with a scanning rate of 15$\arcsec$ per second and a dump time of 1 s. The rotating angle, which represents the cross angle between the 19-beam focal plan array and line of Declination, is set to be 23.4 degrees for smooth super-Nyquist sampling.  

During the observations, a 1.1 K noise from the diode was injected with a period of 2 s, which is synchronized with the sampling rate. Based on an absolute measurement of noise dipole and a factor derived by the difference between noise ON and OFF data for each beam, the observed data were calibrated into antenna temperature T$_A$ in K. Taking the beam width (Full Width at Half Maximum; FWHM) and aperture efficiency $\eta$ from \citet{2020RAA....20...64J}, we convert the antenna temperature (T$_A$) into brightness temperature (T$_B$). The aperture efficiency curves were obtained by repeating observations of the calibrator 3C286 at different zenith angles. 

Using a Gaussian smoothing kernel, the raw data are re-grided and converted to FITS data cube. The beam width is 3$\arcmin$ and the pixel size of the FITS data cube is 1$\arcmin$ $\times$ 1$\arcmin$. The typical system temperature during the observation is about 20 K and the RMS sensitivity of our observation is estimated to be around 70 mK per channel. The pointing of the telescope has an RMS accuracy of 7$.\!\!\arcsec$9. We have compared the results of our data reduction with those of HI4PI \citep{2016A&A...594A.116H}, the HI4PI and FAST  \ion{H} {1} emission profiles agree well with each other and the spectra LSR velocities are consistent.

\subsection{HI4PI Data}

This work makes use of data from all-sky (4$\pi$ sr)  \ion{H} {1} survey \citep{2016A&A...594A.116H}. The HI4PI survey combines \ion{H} {1} data of the Northern hemisphere from EBHIS \citep{2011AN....332..637K,2016A&A...585A..41W} obtained with the 100-m Effelsberg radio telescope with the data of the Southern hemisphere from GASS \citep{2009ApJS..181..398M,2010A&A...521A..17K,2015A&A...578A..78K} taken with the 64-m Parkes radio telescope. The angular and velocity resolutions of the final combined data are 16$.\!\!\arcmin$2 and 1.5 km s$^{-1}$, respectively. The RMS sensitivity of HI4PI \ion{H} {1} data is about 43 mK. The LSR velocity coverage is about $\pm$600 km s$^{-1}$ in the Northern hemisphere and $\pm$480 km s$^{-1}$ in the Southern hemisphere.

\section{Results and discussion}\label{Results}

\begin{figure}[h!]
\centering
\includegraphics[width=0.33\textwidth,angle=0]{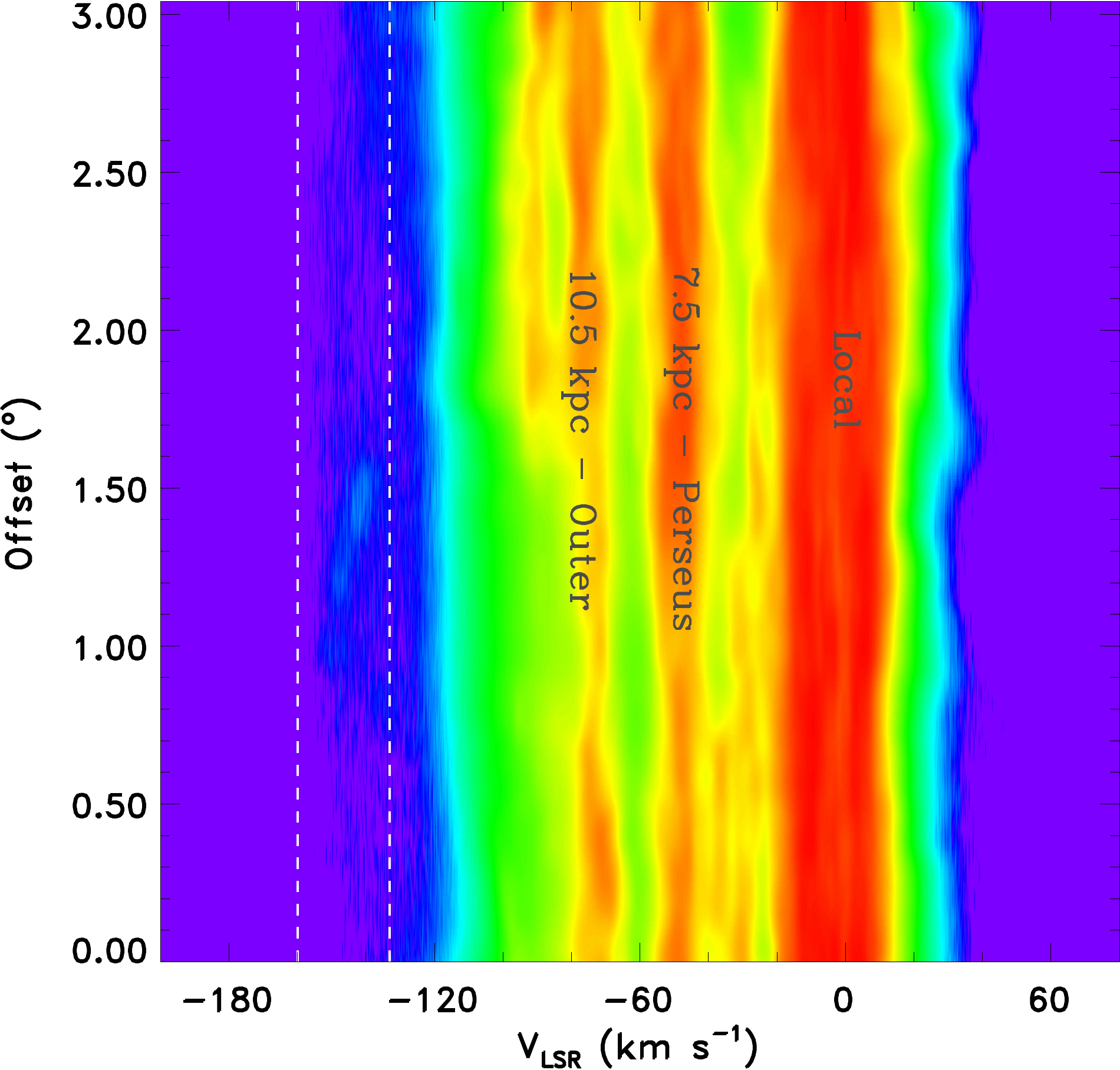}
\includegraphics[width=0.45\textwidth,angle=0]{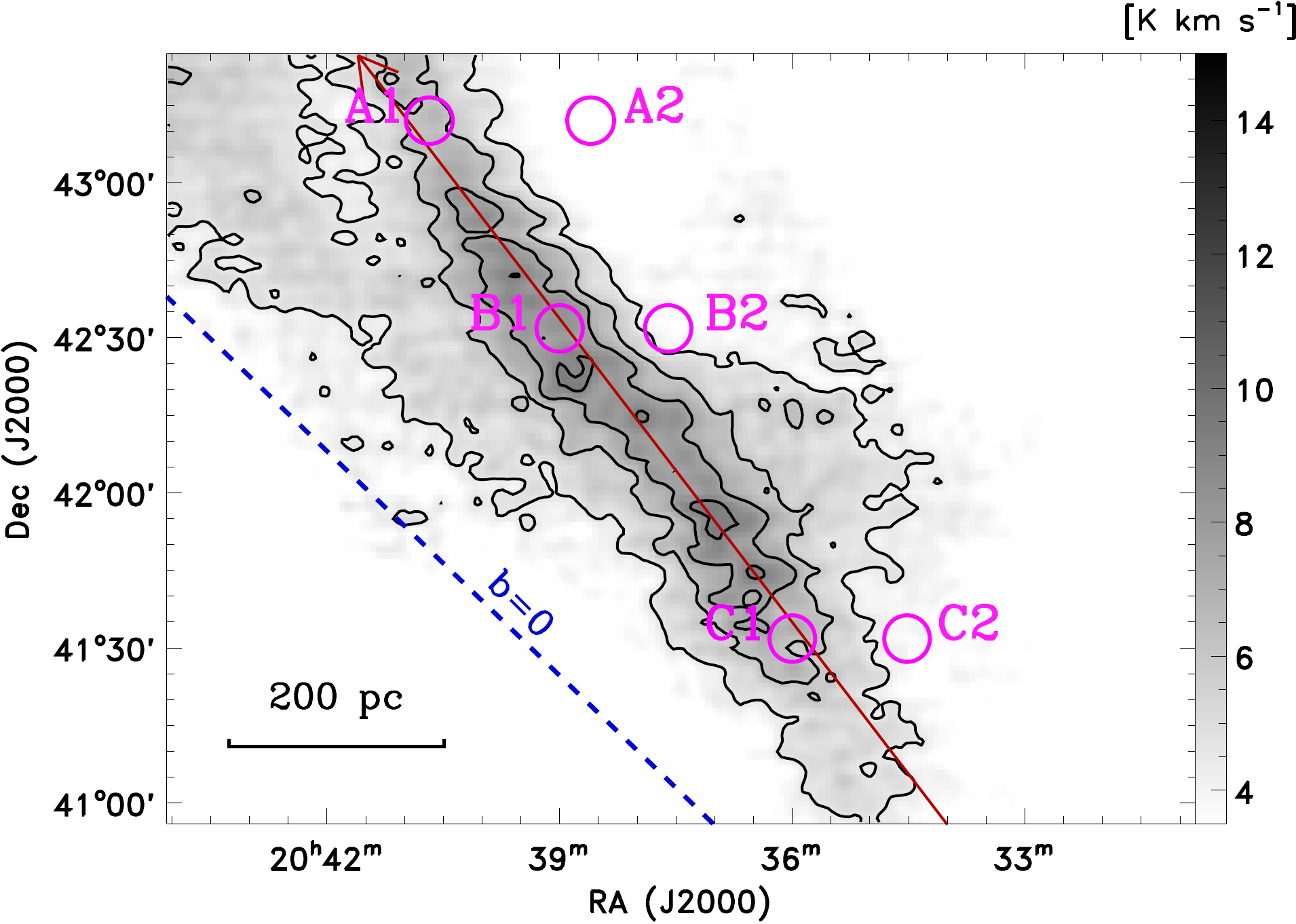}
\caption{Left: position-velocity map of the sky region along the red arrow marked in the right panel. The locations of the spiral arms are derived from \citet{2014ApJ...783..130R,2016ApJ...823...77R}. The dash lines indicate the approximate boundaries of the Extreme Outer Galaxy feature. Right: map of \ion{H} {1} 21-cm emission intensity integrated from $-$160 km s$^{-1}$ to $-$140 km s$^{-1}$. The blue dashed line indicates the IAU-defined mid-plane. The purple circles indicate the positions of the spectra in Figure \ref{fig:dark lane}. The minimal level and interval of the overlaid contours are 4.5 and 1.5 K km s$^{-1}$, respectively.}
\label{pv}
\end{figure}

\begin{figure}[h!]
\centering
\includegraphics[width=0.3\textwidth,angle=0]{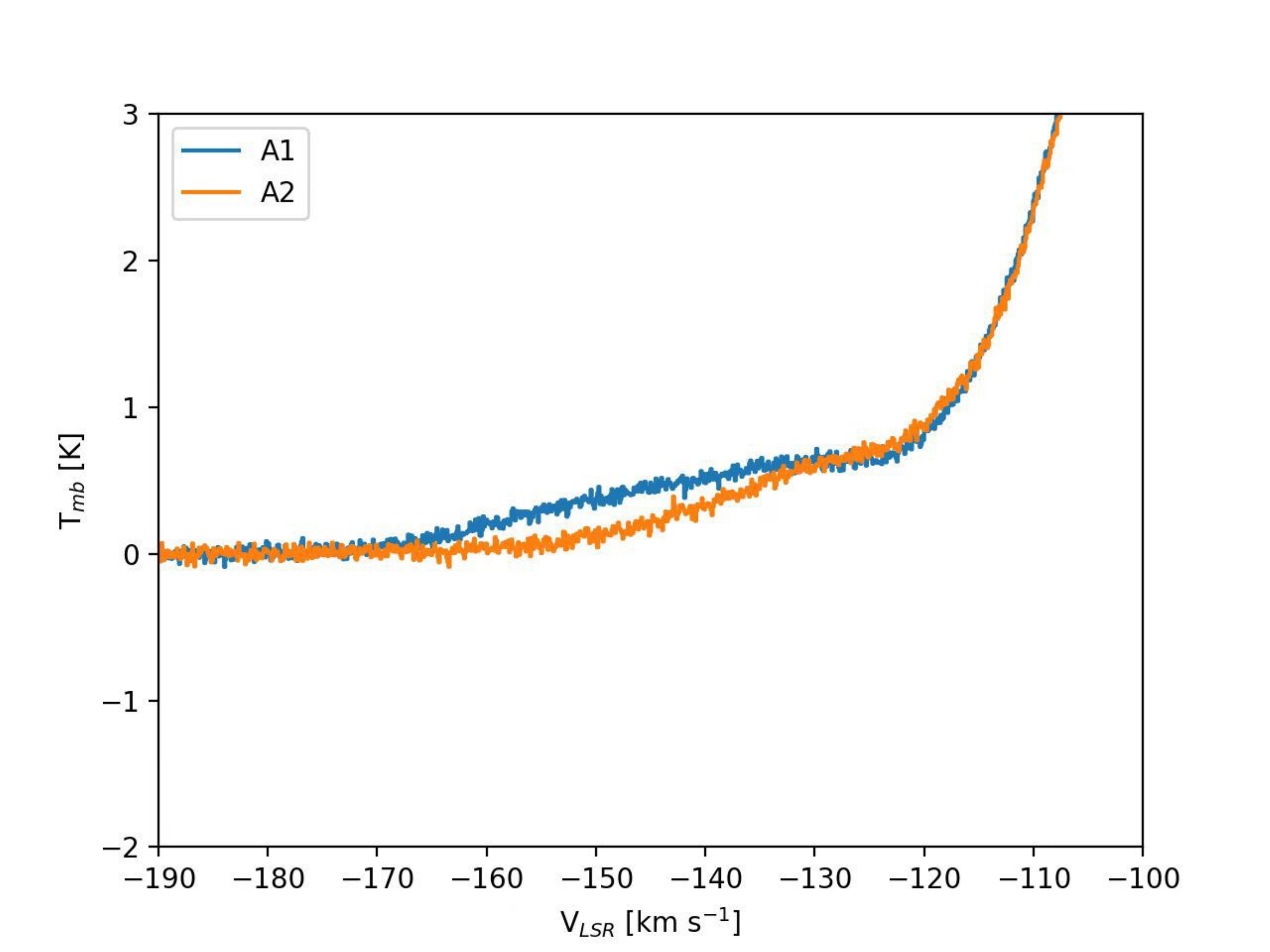}
\includegraphics[width=0.3\textwidth,angle=0]{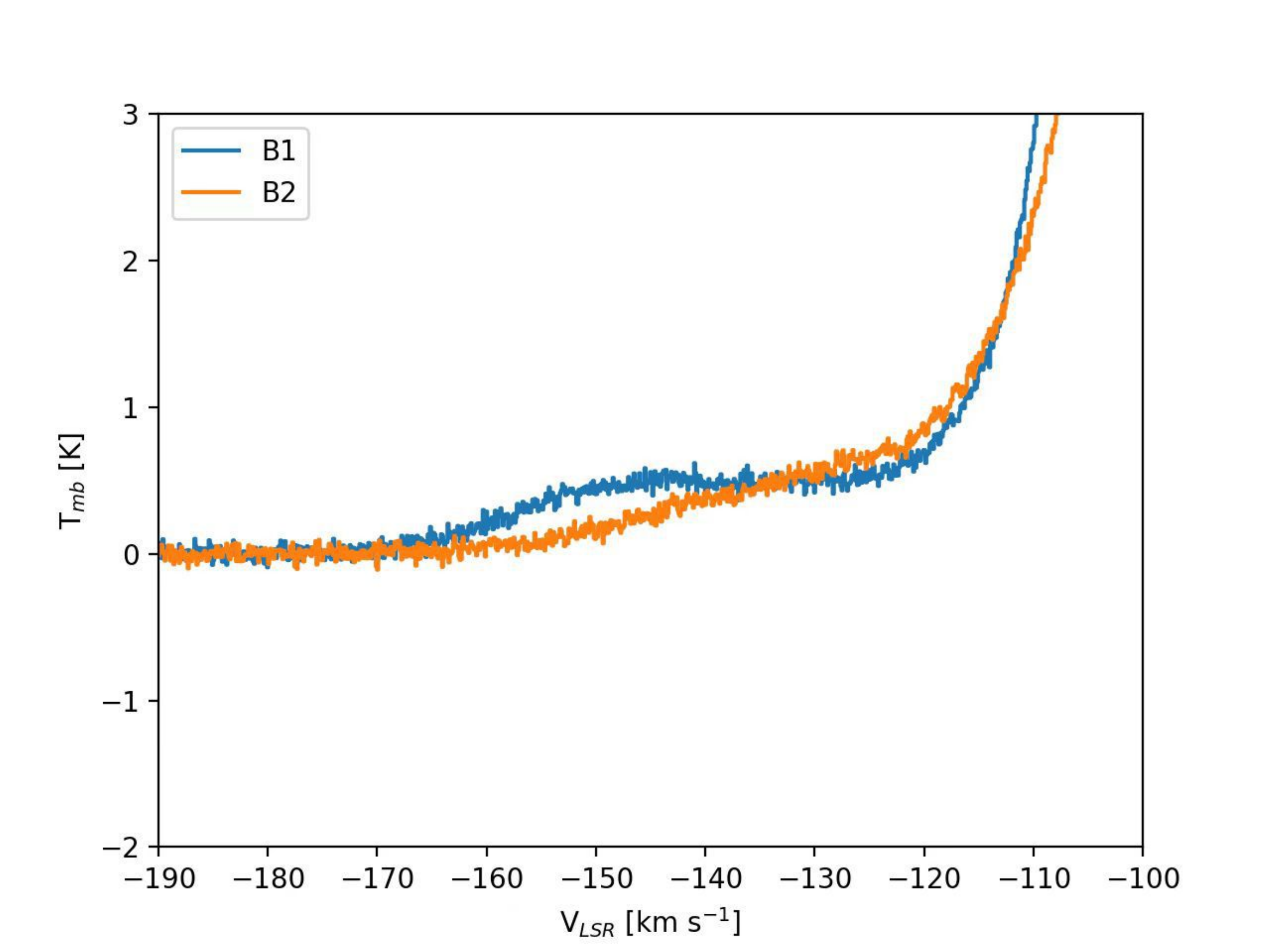}
\includegraphics[width=0.3\textwidth,angle=0]{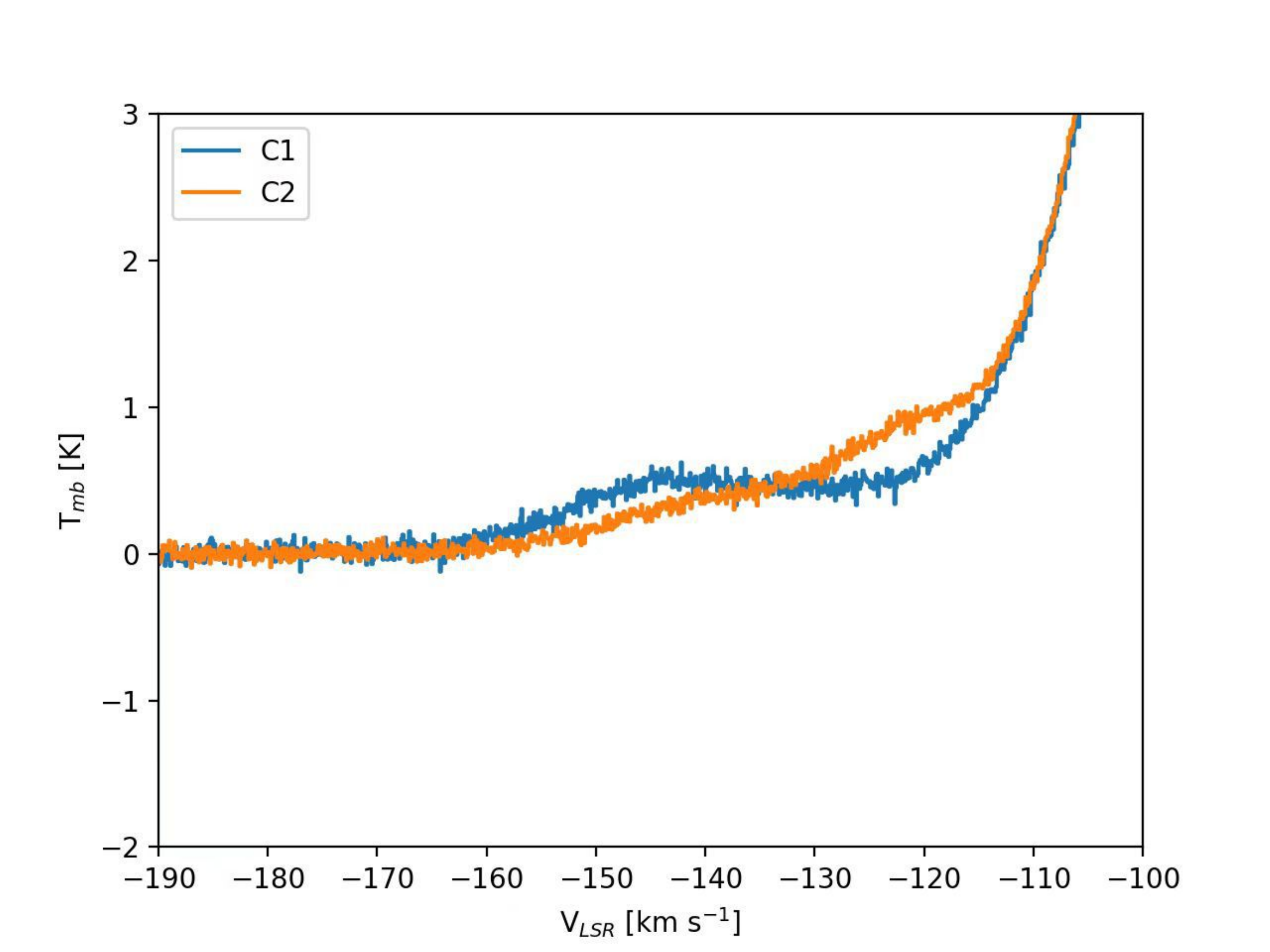}
\caption{FAST \ion{H} {1} spectra averaged over a 0$.\!\!^{\circ}$15 $\times$ 0$.\!\!^{\circ}$15 region taken from the six positions shown in Figure \ref{pv}.}
\label{fig:dark lane}
\end{figure}

The sky region of 20$^{\rm h}$30$^{\rm m}$48$^{\rm s}$ $<$ RA $<$ 20$^{\rm h}$44$^{\rm m}$04$^{\rm s}$ and 40$.\!\!^{\circ}$9 $<$ Dec $<$ 43$.\!\!^{\circ}$4 covers the main part of the Cygnus-X North molecular cloud, which has a velocity range of $-$30 km s$^{-1}$ to 20 km s$^{-1}$ \citep{2010A&A...520A..49S} and is located 1.4 kpc away from the Sun \citep{2012A&A...539A..79R,2013ApJ...769...15X}. In Figure \ref{pv}, however, there remains abundant atomic gas with velocities $<-$30 km s$^{-1}$, implying the presence of \ion{H} {1} gas far behind Cygnus-X. In the velocity range $-$170 km s$^{-1}$ to $-$130 km s$^{-1}$, there appears to be \ion{H} {1} emission not connected to the main velocity components of the Outer arm. The central velocity of the outermost gas is $-$151.1 km s$^{-1}$. According to the rotation model of \citet{2014ApJ...783..130R,2016ApJ...823...77R}, the heliocentric distance of such an \ion{H} {1} cloud is $\sim$21 kpc, corresponding to the distance from the Galactic center of $\sim$22 kpc. The outermost cloud exhibits a filamentary structure and lies approximately parallel to the Galactic mid-plane (b=0, right panel of Figure \ref{pv}). Its length is about 1.1 kpc, much larger than the giant molecular filament Nessie, and well comparable with the large \ion{H} {1} filament Magdalena. If the structure lies at the kinematic distance, it is located far beyond all known giant gas filaments (R$_{gc}$$<$12 kpc), molecular and atomic, in the previous researches \citep{2018ApJ...864..153Z}. Here we name it ``Cattail''. Figure \ref{fig:dark lane} presents the spectra of six positions located within (A1, B1, C1) and immediately outside of Cattail (A2, B2, C2). Compared to A2, B2, C2, the spectra of Cattail (A1, B1, C1) have an additional emission from $-$170 to $-$130 km s$^{-1}$. There is no self-absorption feature in the spectra.

\begin{figure}[h]
\centering
\includegraphics[width=0.8\textwidth,angle=0]{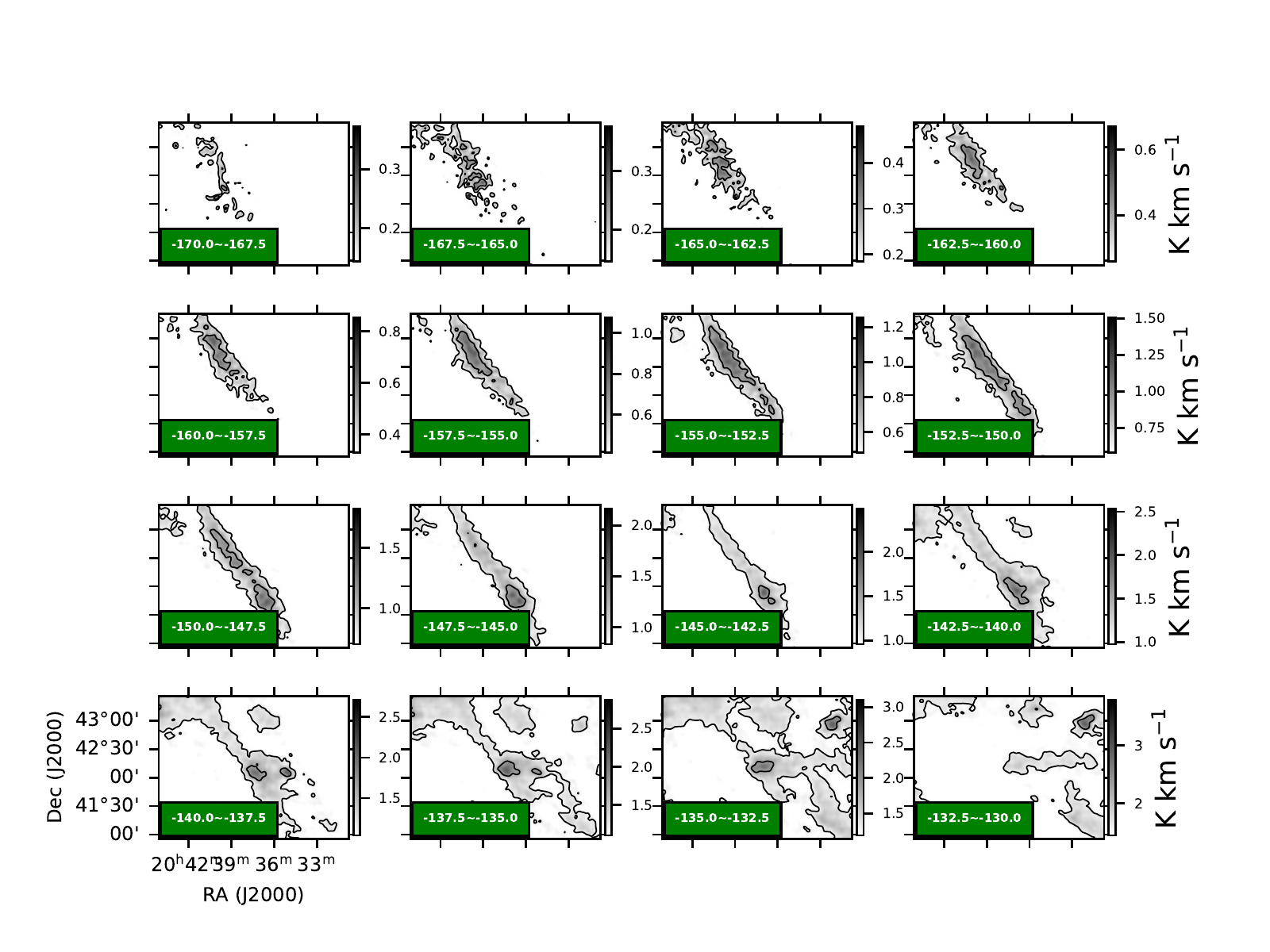}
\caption{Velocity channel map of \ion{H} {1} 21-cm emission of Cattail. The minimal level of the contours is 0.55 times the emission peak and the interval is 0.25 times the peak.}
\label{channel}
\end{figure}

Figure \ref{channel} shows the velocity channel map in \ion{H} {1} 21-cm emission of Cattail. It can be seen that the filament exhibits an overall velocity gradient along its major axis. The southwest end of Cattail has a less negative velocity compared to that of the northeast end.  Most of the atomic gas of the filament has velocities in the range from $-$160 km s$^{-1}$ to $-$140 km s$^{-1}$, with the mean velocity of approximately $-$150 km s$^{-1}$. The intensity distribution and the width of the filament are almost uniform from northeast to southwest. The velocity gradient along the filament is estimated to be 0.02 km s$^{-1}$ pc$^{-1}$.

\begin{figure}[h]
\centering
\includegraphics[width=0.4\textwidth,angle=0]{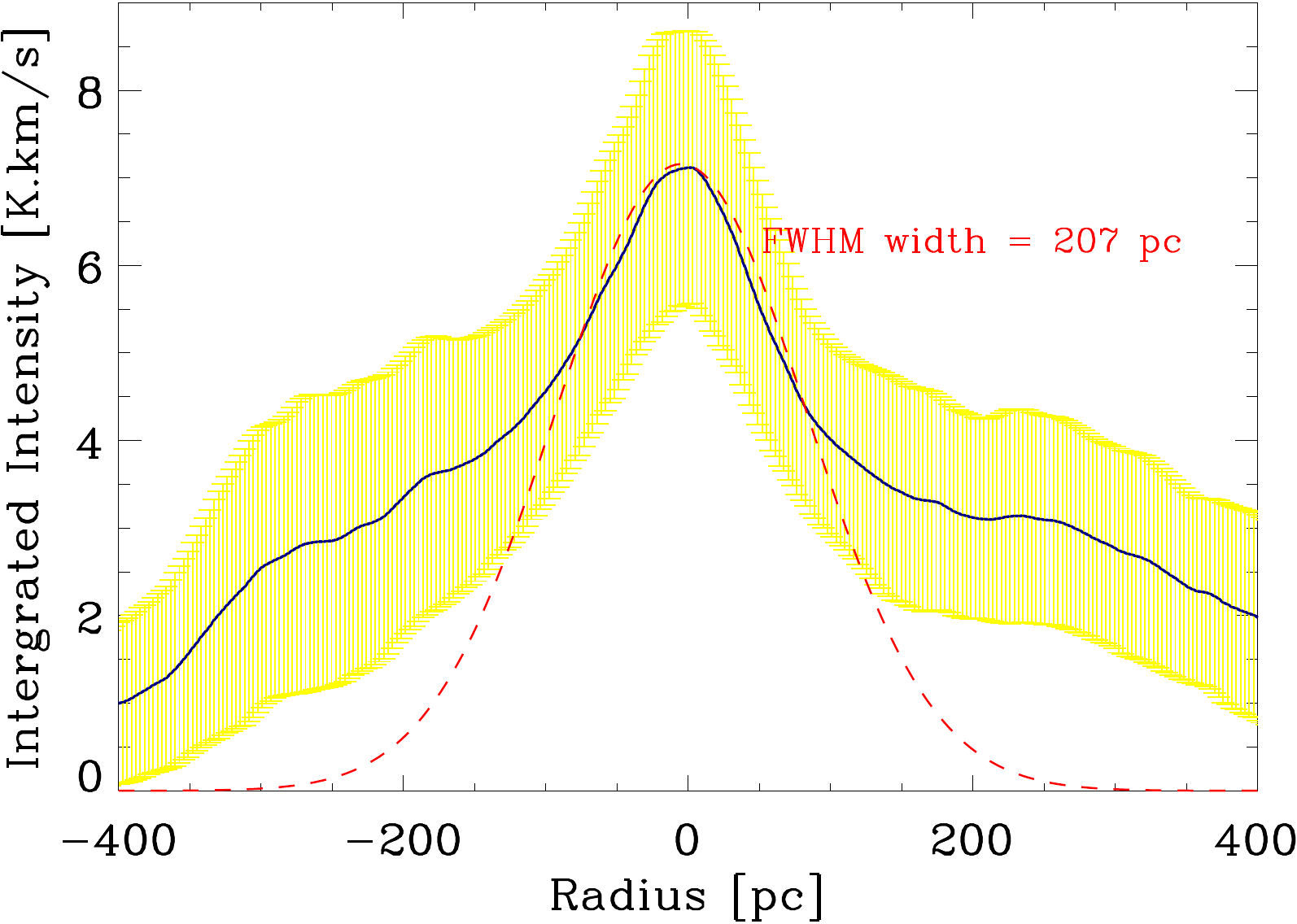}
\caption{Mean radial intensity profile perpendicular to Cattail. The position of the peak intensity is regarded as the center of the profile, i.e., the position of R = 0. The negative and positive distances correspond to the southeast and northwest sides of the filament. The yellow area shows the $\pm1\sigma$ dispersion of the radial profile distribution. The dashed red curve shows the Gaussian fitting to the inner part of the profile.}
\label{width}
\end{figure}

Similar to \citet{2019A&A...623A.142S}, we extract the radial profile of Cattail from the intensity map. Firstly, we derive the direction of the filament spine, which is indicated by a red arrow in the right panel of Figure \ref{pv}. Then, we extract the perpendicular intensity profile for each pixel along the filament spine. Finally, we average the profiles of all pixels along the filament and apply Gaussian fitting to the mean radial profile. In addition to the emission of Cattail, there is residual emission in the sky region which comes from the velocity distribution of the closer arms (Figure \ref{fig:dark lane}). In this case, the calculated width would increase with the fitting range \citep{2014MNRAS.445.2900S,2017MNRAS.466.2529P}. To be less influenced by this bias, we apply Gaussian fitting to the inner part of the radial profile. The intensity radial profile of Cattail is almost symmetrical (Figure \ref{width}). The width (FWHM) of the filament is 207 pc, which means that the aspect ratio of Cattail is about 5:1. Given that the width is comparable to the scale height of gas in a spiral arm \citep{2009ARA&A..47...27K}, Cattail could also be a density enhancement along a spiral arm, we will come back to this point later.

Considering that the peak antenna temperature is about 1 K (Figure \ref{fig:dark lane}) and there is no \ion{H} {1} narrow self-absorption feature in the spectra (see Figure \ref{fig:dark lane}), the emission should be optically thin and the column density can be calculated according to the following formula \citep{2009tra..book.....W}

\begin{equation}
\label{equa}
\rm N_{HI}\ [cm^{-2}] = 1.823 \times 10^{18} \int T_B(v)dv\ [K\ km\ s^{-1}]
\end{equation}

We calculated the mass of Cattail to be 6.5 $\times$ 10$^4$ M$_{\odot}$. The linear mass density is 60 M$_{\odot}$ pc$^{-1}$. The peak column density of Cattail is 1.7 $\times$ 10$^{19}$ cm$^{-2}$ (0.1 M$_{\odot}$ pc$^{-2}$), which is close to the median value of \ion{H} {1} filaments with latitudes $|$b$|$ $>$ 20$^{\circ}$ \citep{2016ApJ...821..117K} and lower than the threshold (N $\ge$ 1 $\times$ 10$^{22}$ cm$^{-2}$) for high column density gas adopted by \citet{2018ApJ...864..153Z}. Assuming the depth to be the same as the width (207 pc), the peak volume density of the filament is calculated to be 0.03 cm$^{-3}$, which is close to the average Galactic mid-plane volume density at $R_{gc}$=22 kpc (0.01 cm$^{-3}$) \citep{2009ARA&A..47...27K}. The physical parameters of Cattail are listed in Table \ref{tab1}.

Both simulations \citep{1992ApJ...388..392I,1997ApJ...480..681I} and observations \citep{2010A&A...518L.102A,2010A&A...518L.103M,2012A&A...540L..11S,2013A&A...551C...1S} have shown that filaments tend to be unstable to radial collapse if the line mass is greater than the critical value defined as $M_{\rm vir} = 2\sigma_{\rm tot}^{2}/G$ \citep{2000MNRAS.311...85F}. The total velocity dispersion $\sigma_{\rm tot}$ can be obtained by the FWHM line width ($\Delta v/\sqrt{8 \rm ln 2}$), which is 12 km s$^{-1}$ for Cattail. \ion{H} {1} filaments are normally cold with a typical excitation temperature T$_{ex}$$\sim$50 K \citep{2020A&A...639A..26K}, corresponding to an isothermal sound speed of approximately 0.65 km s$^{-1}$. Therefore, the turbulence may be supersonic in Cattail. The critical line mass M$_{vir}$ is calculated to be 1.1$\times$ 10$^4$ $M_\sun$ pc$^{-1}$, which is much larger than the linear mass density M$_{line}$ (60 M$_{\odot}$ pc$^{-1}$). Thus, Cattail is gravitationally unbound and may be in an expanding state, unless confined by an external pressure \citep{2012A&A...542A..77F}.

\begin{deluxetable}{lcccrrcccrrc}
\decimals
\tabletypesize{\footnotesize}
\tablewidth{0pt}
\tablenum{1}
\tablecaption{Physical parameters of Cattail\label{tab1}}
\tablehead{
\colhead{parameter} & \colhead{value} \\
\colhead{ }  & \colhead{ }   }
\startdata
Galactic longitude: l & 81$.\!\!^{\circ}$49     \\
Galactic latitude: b & 0$.\!\!^{\circ}$65    \\
Right ascension: $\alpha$J2000 & 20h37m54s     \\
Declination: $\delta$J2000 & +42d14m40s    \\
Center velocity: & $-$151.1 km/s    \\
Kinematic distance: D & 21 kpc     \\
Galactocentric radius: R$_{gc}$ & 22 kpc   \\
Length & 1.1 kpc    \\
Width (FWHM) & 207 pc    \\
Aspect ratio & 5:1    \\
Mass &  6.5 $\times$ 10$^4$ M$_{\odot}$   \\
Linear Mass &   60 M$_{\odot}$ pc$^{-1}$  \\
Line width: $\rm{\Delta V}$ & 12 km/s    \\
Peak column density & 1.7 $\times$ 10$^{19}$ cm$^{-2}$ (0.1 M$_{\odot}$ pc$^{-2}$)   \\
Peak volume density &   0.03 cm$^{-3}$ \\
\enddata
\tablecomments{Rows 1-5 give the central positions in the PPV space. The kinematic distance is derived according to the Galactic rotation model A5 of \citet{2014ApJ...783..130R}. The volume density is defined as N$_{HI}$/depth, where depth is assumed to be the same as width.}
\end{deluxetable}

A velocity range of $|V_{LSR}|$ $>$ 90 km s$^{-1}$ is generally used to separate high-velocity clouds (HVC) from gas at low and intermediate velocities. However, this operational definition is insufficient considering that the velocities of some clouds can be well understood in terms of Galactic rotation but exceed 90 km s$^{-1}$ \citep{2016ApJ...823...77R}. Similarly with \citet{1991A&A...250..499W}, we adopt a ``forbidden velocity'' of $|V_{LSR}- V_{rot}|$ $>$ 50 km s$^{-1}$ to identify HVCs, where $V_{rot}$ is the rotation curve of the Milky Way. In the direction of Cattail, the filament can be a HVC only when its heliocentric distance is within 12.5 kpc (R$_{gc} < 14$ kpc). However, the Galactic Latitude of Cattail is 0$.\!\!^{\circ}$65, which would put it within the flaring FWHM of the disk when R$_{gc} < 14$ kpc \citep{2009ARA&A..47...27K}. As HVCs travel through hot halo gas, they would be heated and photon ionized, which causes thermal instabilities between the skin and the inner region of the clouds \citep{2012ARA&A..50..491P}. Moreover, the headwind pressures can induce dynamical instabilities, e.g., shear-driven disturbance in HVCs, which makes them fragment into head-tail clouds \citep{2004ASSL..312..313B} and increase their line widths \citep{2011MNRAS.418.1575P}. The \ion{H} {1} 21-cm line width of Cattail is 12 km s$^{-1}$, significantly smaller than the typical value of 20-30 km s$^{-1}$ for HVCs \citep{2003ApJ...591L..33L,2017ARA&A..55..389T,2020ApJ...902..154B}. The cloud does not seem to be fragmenting. Also, Cattail has a filamentary morphology parallel to the Galactic mid-plane with a mass less than 10$^5$ M$_{\odot}$, and if it is a HVC, it would have been fully ionized during passing through the Galaxy halo \citep{2007ApJ...670L.109B,2009ApJ...698.1485H,2011ApJ...739...30K}. Therefore, the properties of Cattail do not appear to be compatible with that of a HVC.

Assuming that the velocity of Cattail is attributed to the Galactic rotation, the structure is located beyond all the known spiral arms in the first quadrant. We then examine other possible origins of the extreme velocity of the Cattail. We have discussed above that Cattail does not seem to be an HVC. Combining the results of \citet{2014ApJS..212....1A,2014BASI...42...47G}, there are no SNRs or H II regions with regular sizes $>$2$.\!\!^{\circ}$5 near Cattail. There is no dwarf satellite galaxy near Cattail either \citep{2012AJ....144....4M}. Hence, there is no evidence for Cattail's velocity coming from the influences of stellar feedback or a tidal stream, and the most likely explanation is the Galactic rotation. Moreover, taking the column density and the depth of Cattail to be 1.7 $\times$ 10$^{19}$ cm$^{-2}$ and $\sim$0$.\!\!^{\circ}$5, respectively, the volume density of Cattail would be significantly lower than the typical value of the Milky Way when it is located at R$_{gc}$ $<$ 15 kpc \citep{2009ARA&A..47...27K}. Both a filament and a spiral arm should have a large density contrast compared to the surrounding medium. Therefore, we consider that Cattail is located at a heliocentric distance and Galactocentric distance of both $\sim$20 kpc. 

\begin{figure}[h]
\centering
\includegraphics[width=0.4\textwidth,angle=0]{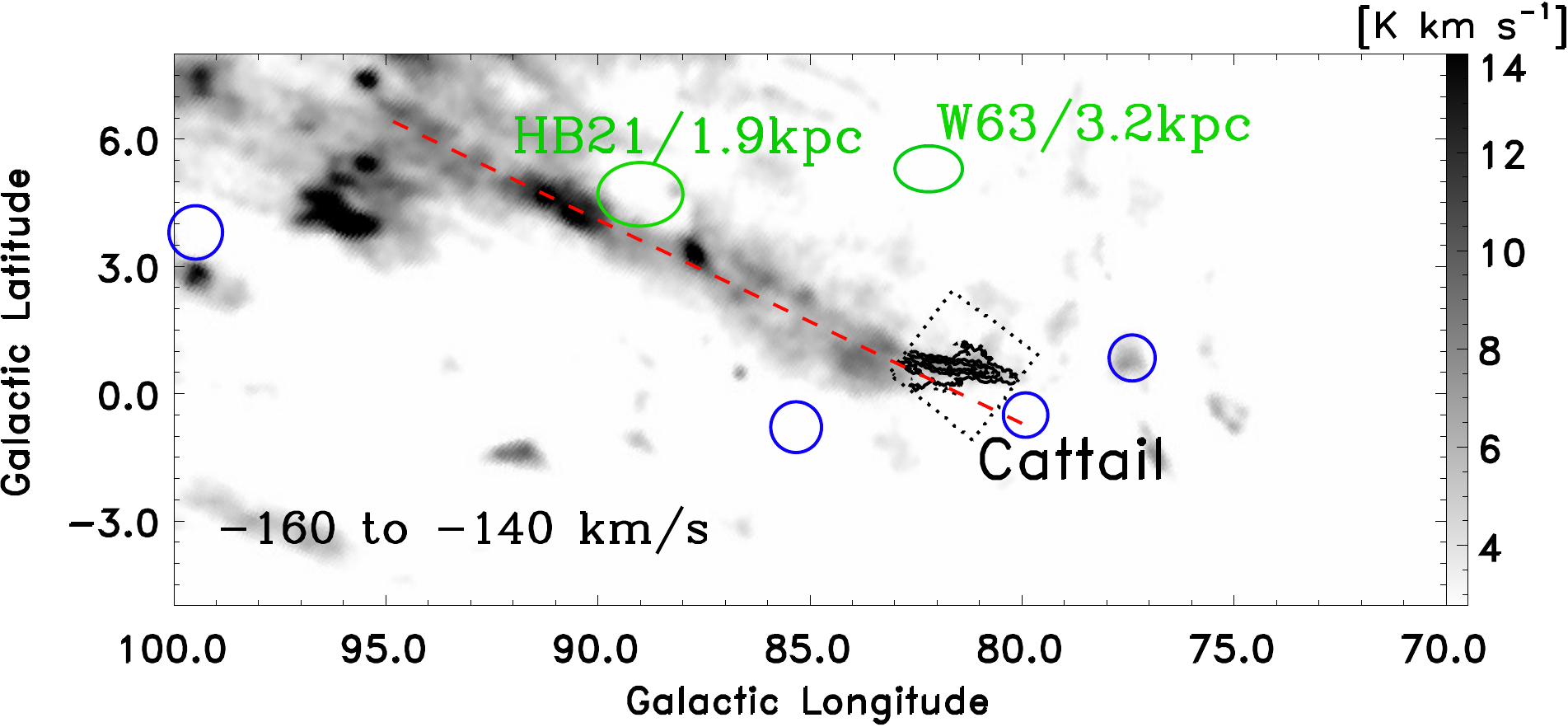}
\includegraphics[width=0.4\textwidth,angle=0]{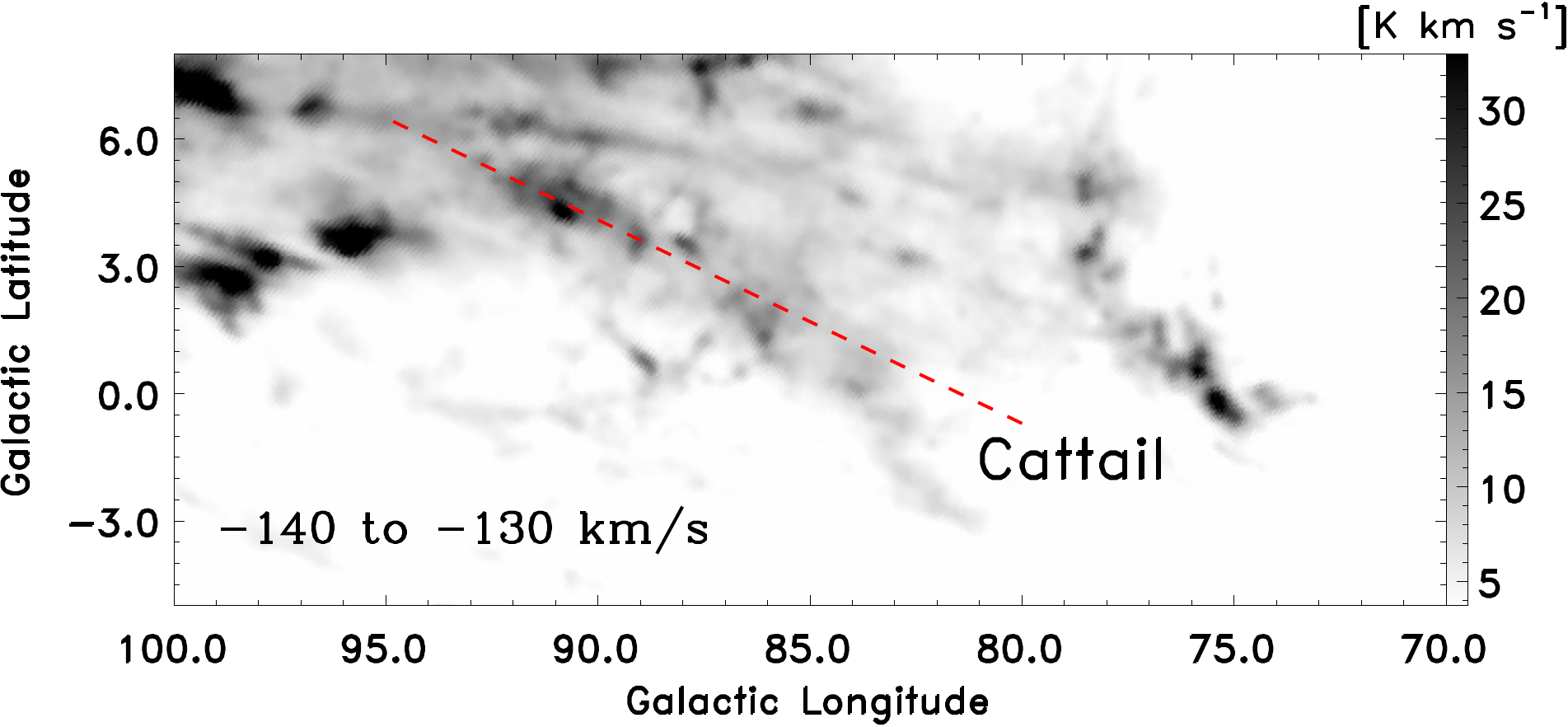}
\includegraphics[width=0.4\textwidth,angle=0]{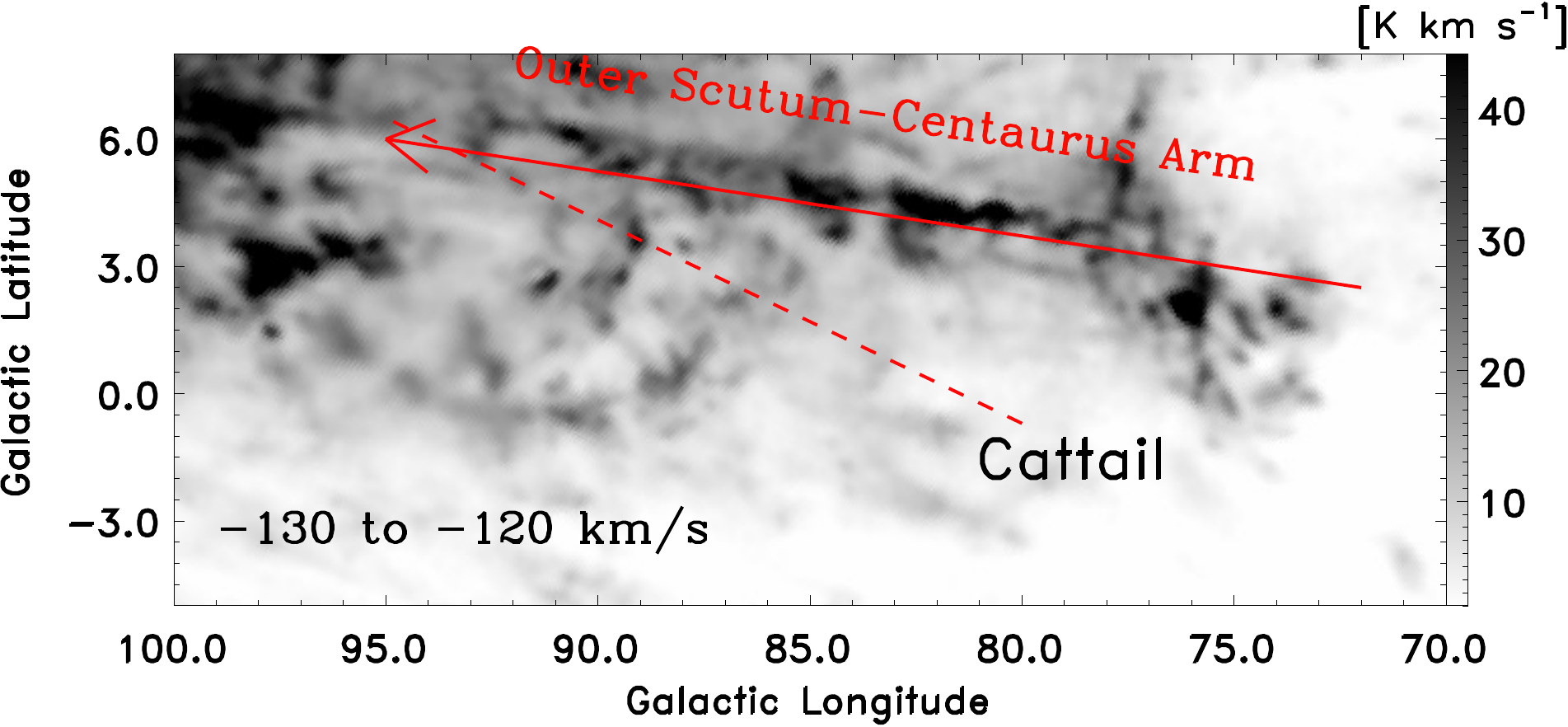}
\includegraphics[width=0.4\textwidth,angle=0]{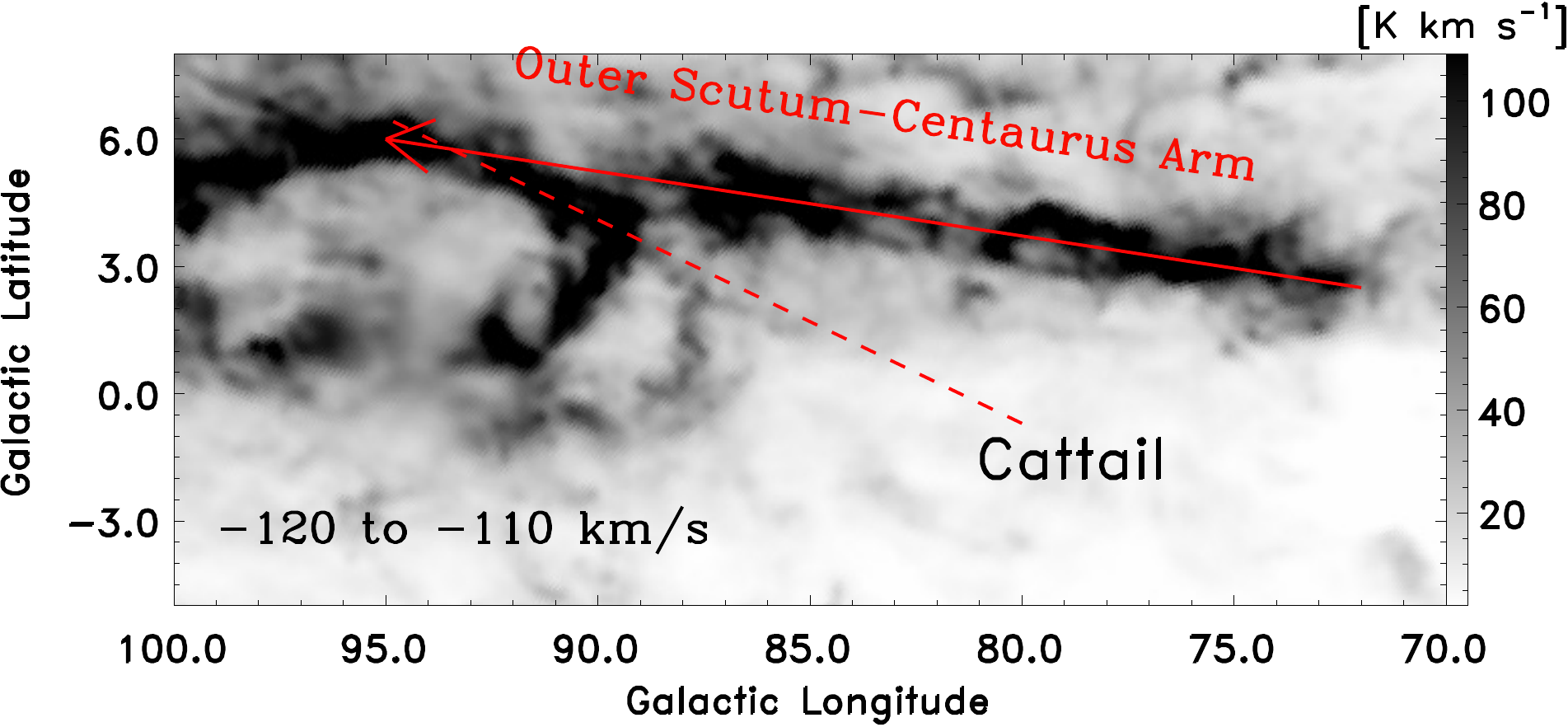}
\caption{H I integrated intensity maps of the HI4PI data in the velocity ranges from $-$160 km s$^{-1}$ to $-$140 km s$^{-1}$, $-$140 km s$^{-1}$ to $-$130 km s$^{-1}$, $-$130 km s$^{-1}$ to $-$120 km s$^{-1}$, and $-$120 km s$^{-1}$ to $-$110 km s$^{-1}$, respectively. The dashed box in the top left panel indicates the survey region of FAST observation. The green and blue ellipses indicate the locations of the Galactic supernova remnants and \ion{H} {2} regions with the angular sizes $> 1^{\circ}$. The red arrows indicate the path of position-velocity plot in Figure \ref{pv osc}.}
\label{hi4pi}
\end{figure}

Using the HI4PI data, we have investigated the complete morphology of Cattail (Figure \ref{hi4pi}). Cattail exhibits an even much larger dimension while retaining a coherent velocity of $-$160 km s$^{-1}$ to $-$140 km s$^{-1}$. The FAST observations cover the southwest end of the full structure while the northeast end is located at about $l$ $=$ 93$^{\circ}$ and $b$ $=$ 5$.\!\!^{\circ}$5. We designate them as the FAST end and HI4PI end in this work, respectively. The full length of the complete Cattail reachs $\sim$15$^{\circ}$ (5 kpc). 

\begin{figure}[h]
\centering
\includegraphics[width=0.35\textwidth,angle=0]{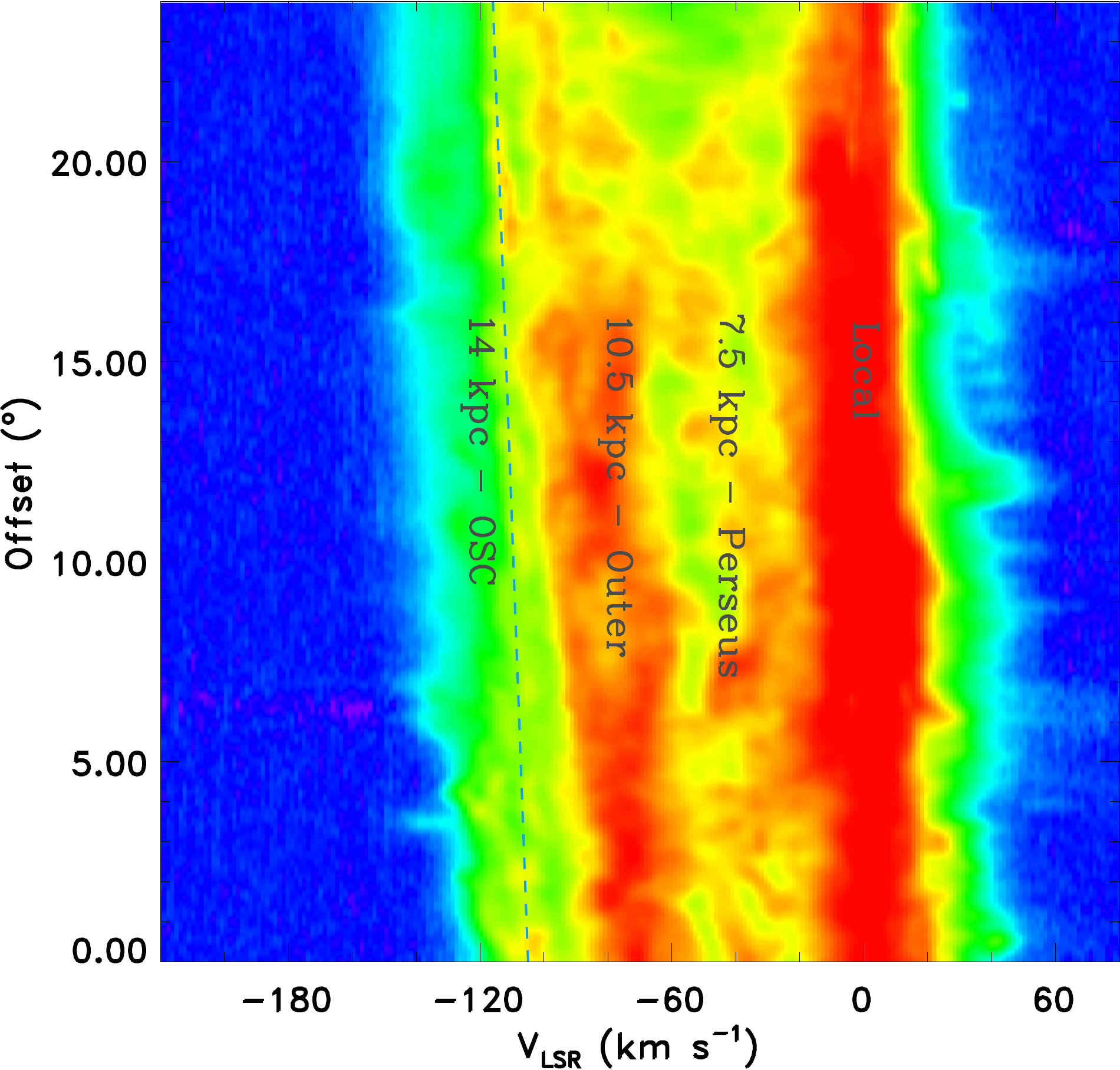}
\caption{Position-velocity map along the arrow marked in Figure \ref{hi4pi}.}
\label{pv osc}
\end{figure}

In the velocity range from $-$130 km s$^{-1}$ to $-$110 km s$^{-1}$, there is another new elongated structure between 70$^{\circ}$ $<$ $l$ $<$ 100$^{\circ}$. From its morphology (Figure \ref{hi4pi}) and velocity (Figure \ref{pv osc}), it is apparently the extension of the OSC arm previously only seen at 0$^{\circ}$ $<$ $l$ $<$ 70$^{\circ}$ \citep{2011ApJ...734L..24D}. The full structure of the OSC in the first Galactic quadrant is then seen for the first time. In projection, Cattail is connected to the OSC at about $l=93^{\circ}$. However, Cattail does not appear to be physically connected to the OSC arm. First, the velocity of Cattail is different from that of the OSC by 30 km s$^{-1}$. Second, if Cattail is connected to the OSC at a Galactic Longitude of about $93^{\circ}$ and extends to about $80^{\circ}$, it means that Cattail joins the OSC arm at the downstream end with respect to the Galactic rotation and leaves the arm toward the upstream end, i.e., in a pattern opposite to the inertial direction of the Galactic rotation; such a structure is inconsistent with the picture of a spur originating from a spiral arm (e.g., \citealp{2006MNRAS.367..873D}). Thus, combining the large scale map of HI4PI, Cattail may have a much larger dimension but still appears to be located well above the OSC.

\begin{figure}[h]
\centering
\includegraphics[width=0.4\textwidth,angle=0]{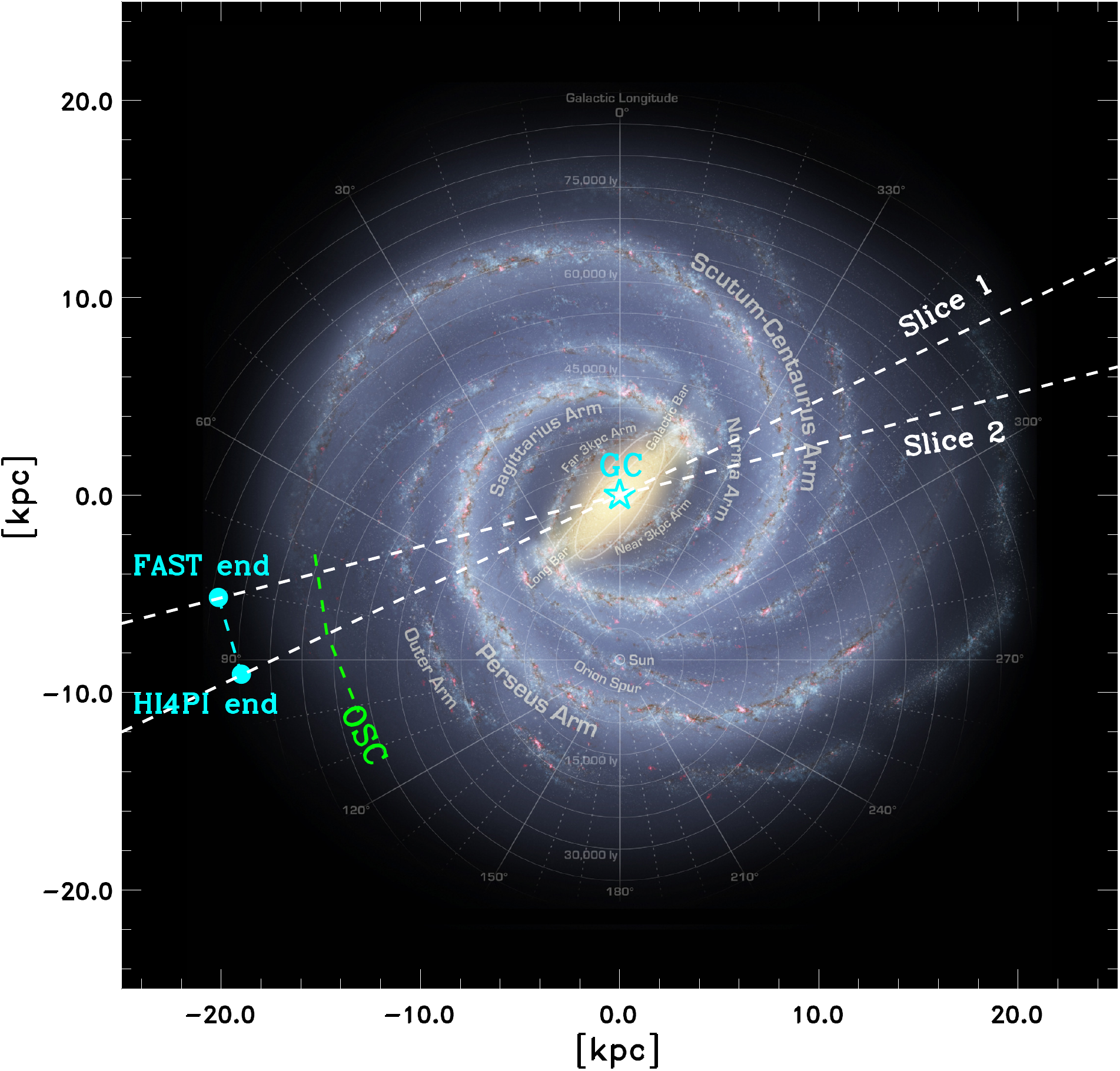}\\
~~~~~~~~~~~~~~~~\includegraphics[width=0.8\textwidth,angle=0]{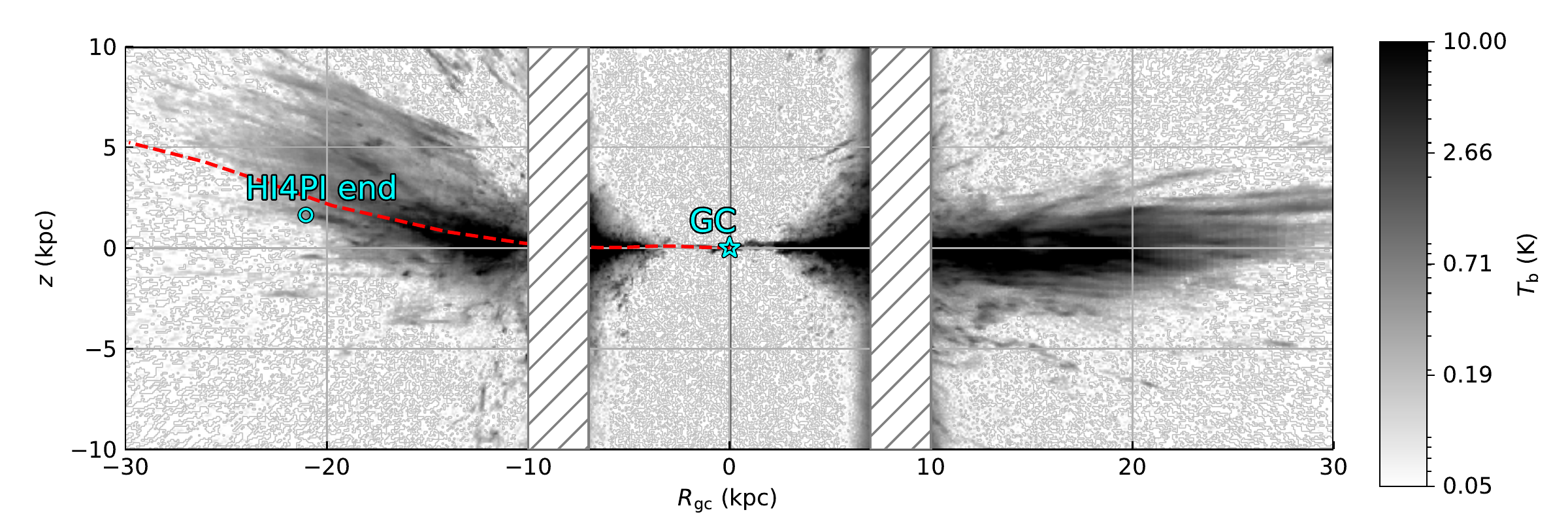}\\
~~~~~~~~~~~~~~~~\includegraphics[width=0.8\textwidth,angle=0]{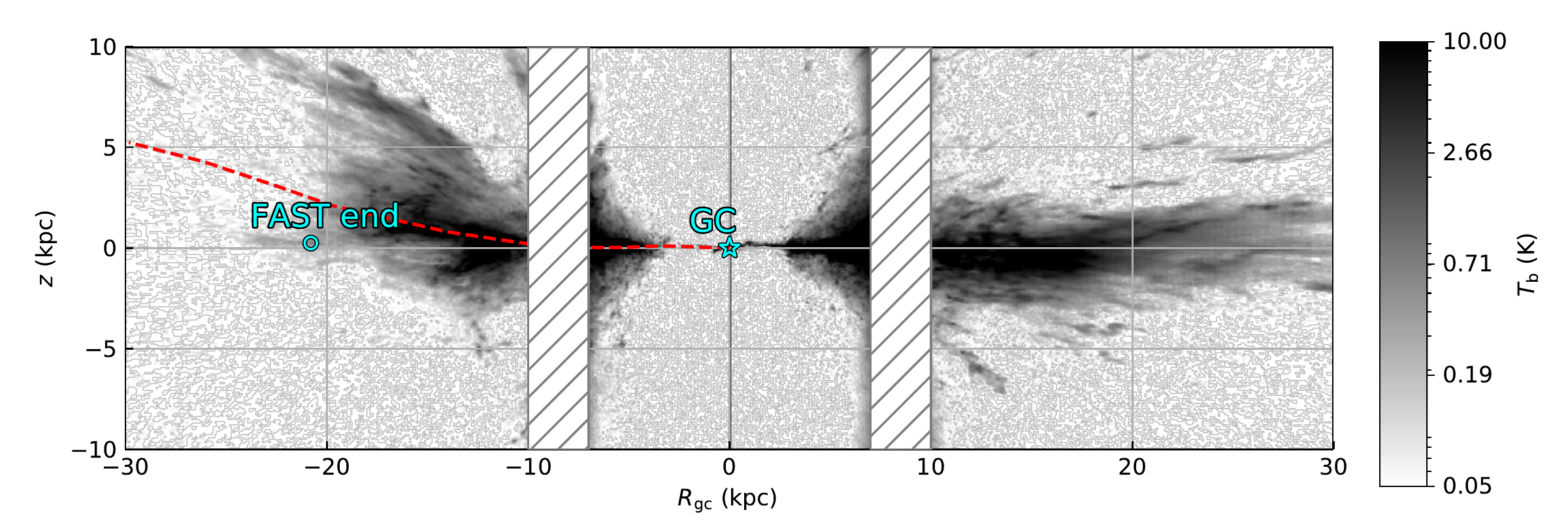}
\caption{Top: artist's conception view of the Milky Way (R. Hurt: NASA/JPL-Caltech/SSC). The new part of the OSC and the Cattail identified in this work are indicated with the green and blue dashed line, respectively. Middle: intensity distribution along slice 1 in the top panel, which derived from the HI4PI data under the Galactic rotation model A5 of \citet{2014ApJ...783..130R}. The red dashed line indicates a sketch of the warped plane near Cattail derived from \citet{2006ApJ...643..881L,2009ARA&A..47...27K}. Bottom: intensity distribution along slice 2 in the top panel.}
\label{galaxy}
\end{figure}

Based on the above analysis, we suggest two possible explanations for Cattail: it is a giant filament with a length of $\sim$5 kpc, or part of a new arm in the EOG (Figure \ref{galaxy}). Only until recently, simulations of giant filaments appeared in the literature. As the numerical simulations show, the shear from Galactic rotation plays a critical role in the formation of large-scale filaments \citep{2014MNRAS.441.1628S,2016MNRAS.458.3667D}. Since the shearing motion tends to stretch out and align gas with spiral arms, giant filaments tend to form in spiral arms and interarm regions. Unlike most of the giant molecular filaments which are associated with spiral arms and lie within 30 pc from the physical Galactic mid-plane (Wang et al. 2016; Zucker et al. 2018), Cattail is far beyond the outermost arm OSC. Figure \ref{galaxy} shows that the Galactic disk is significantly warped in the first quadrant of the EOG, which is in line with the forecasts from \citet{2006ApJ...643..881L,2007A&A...469..511K,2009ARA&A..47...27K}.  The $Z$ scale height of the HI4PI end is about 2 kpc, which approximates the warped scale of the physical Galactic mid-plane at that Galactocentric distance. Comparatively, the FAST end of Cattail is located at the Galactic Latitude of 0$.\!\!^{\circ}$65, corresponding to about 200 pc upon the IAU defined mid-plane ($b=0^{\circ}$), which is flat from the Galactic center to the edge of the Galaxy. Thus, the complete Cattail may originate from the warped physical Galactic mid-plane at a longitude of $\sim$$93^{\circ}$ and extend to a longitude of $\sim$$80^{\circ}$ with $Z \sim$ 200 pc. If Cattail is a gas filament located beyond the OSC, how is such a huge structure formed? Alternatively, if Cattail is part of a new spiral arm in the EOG, it is also puzzling that the new arm does not fully follow the Galactic warp, given that the FAST end is about 1.8 kpc away from the warped Galactic disk mid-plane. While these questions remain open with the existing data, the observations provide new insights into our understanding of the Galactic structure.

\section{Acknowledgements}

This work is supported by National Key R$\&$D Program of China No. 2017YFA0402600 and the National Natural Science Foundation of China (NSFC) grant U1731237. C.L. acknowledges the supports by China Postdoctoral Science Foundation No. 2021M691532 and Jiangsu Postdoctoral Research Funding Program No. 2021K179B. Y.C. is partially supported by the Scholarship No. 201906190105 of the China Scholarship Council and the Predoctoral Program of the Smithsonian Astrophysical Observatory (SAO). We thank the referee for his/her thoughtful comments which improved this paper. We would like to thank the FAST staff for their supports during the observation and thank Junzhi Wang, Zhiyu Zhang, and Bing Liu for the helpful discussions on data reduction. This work makes use of publicly released data from the HI4PI survey which combines the EBHIS in the Northern hemisphere with the GASS in the Southern hemisphere. EBHIS is based on observations with the 100-m telescope of the MPIfR (Max-Planck-Institut fur Radioastronomie) at Effelsberg. The Parkes Radio Telescope is part of the Australia Telescope which is funded by the Commonwealth of Australia for operation as a National Facility managed by CSIRO.



\bibliographystyle{apj}
\bibliography{my_ref_all}

\begin{thebibliography}{68}
\expandafter\ifx\csname natexlab\endcsname\relax\def\natexlab#1{#1}\fi

\bibitem[{{Abreu-Vicente} {et~al.}(2016){Abreu-Vicente}, {Ragan},
  {Kainulainen}, {Henning}, {Beuther}, \& {Johnston}}]{2016A&A...590A.131A}
{Abreu-Vicente}, J., {Ragan}, S., {Kainulainen}, J., {et~al.} 2016, \aap, 590,
  A131

\bibitem[{{Anderson} {et~al.}(2014){Anderson}, {Bania}, {Balser}, {Cunningham},
  {Wenger}, {Johnstone}, \& {Armentrout}}]{2014ApJS..212....1A}
{Anderson}, L.~D., {Bania}, T.~M., {Balser}, D.~S., {et~al.} 2014, \apjs, 212,
  1

\bibitem[{{Andr{\'e}} {et~al.}(2014){Andr{\'e}}, {Di Francesco},
  {Ward-Thompson}, {Inutsuka}, {Pudritz}, \& {Pineda}}]{2014prpl.conf...27A}
{Andr{\'e}}, P., {Di Francesco}, J., {Ward-Thompson}, D., {et~al.} 2014,
  Protostars and Planets VI, 27

\bibitem[{{Andr{\'e}} {et~al.}(2010){Andr{\'e}}, {Men'shchikov}, {Bontemps},
  {K{\"o}nyves}, {Motte}, {Schneider}, {Didelon}, {Minier}, {Saraceno},
  {Ward-Thompson}, {di Francesco}, {White}, {Molinari}, {Testi}, {Abergel},
  {Griffin}, {Henning}, {Royer}, {Mer{\'{\i}}n}, {Vavrek}, {Attard},
  {Arzoumanian}, {Wilson}, {Ade}, {Aussel}, {Baluteau}, {Benedettini},
  {Bernard}, {Blommaert}, {Cambr{\'e}sy}, {Cox}, {di Giorgio}, {Hargrave},
  {Hennemann}, {Huang}, {Kirk}, {Krause}, {Launhardt}, {Leeks}, {Le Pennec},
  {Li}, {Martin}, {Maury}, {Olofsson}, {Omont}, {Peretto}, {Pezzuto}, {Prusti},
  {Roussel}, {Russeil}, {Sauvage}, {Sibthorpe}, {Sicilia-Aguilar}, {Spinoglio},
  {Waelkens}, {Woodcraft}, \& {Zavagno}}]{2010A&A...518L.102A}
{Andr{\'e}}, P., {Men'shchikov}, A., {Bontemps}, S., {et~al.} 2010, \aap, 518,
  L102

\bibitem[{{Barger} {et~al.}(2020){Barger}, {Nidever}, {Huey-You}, {Lehner},
  {Rueff}, {Freeman}, {Birdwell}, {Wakker}, {Bland-Hawthorn}, {Benjamin}, \&
  {Ciampa}}]{2020ApJ...902..154B}
{Barger}, K.~A., {Nidever}, D.~L., {Huey-You}, C., {et~al.} 2020, \apj, 902,
  154

\bibitem[{{Beuther} {et~al.}(2016){Beuther}, {Bihr}, {Rugel}, {Johnston},
  {Wang}, {Walter}, {Brunthaler}, {Walsh}, {Ott}, {Stil}, {Henning},
  {Schierhuber}, {Kainulainen}, {Heyer}, {Goldsmith}, {Anderson}, {Longmore},
  {Klessen}, {Glover}, {Urquhart}, {Plume}, {Ragan}, {Schneider},
  {McClure-Griffiths}, {Menten}, {Smith}, {Roy}, {Shanahan}, {Nguyen-Luong}, \&
  {Bigiel}}]{2016A&A...595A..32B}
{Beuther}, H., {Bihr}, S., {Rugel}, M., {et~al.} 2016, \aap, 595, A32

\bibitem[{{Bland-Hawthorn} {et~al.}(2007){Bland-Hawthorn}, {Sutherland},
  {Agertz}, \& {Moore}}]{2007ApJ...670L.109B}
{Bland-Hawthorn}, J., {Sutherland}, R., {Agertz}, O., \& {Moore}, B. 2007,
  \apjl, 670, L109

\bibitem[{{Butler Burton} {et~al.}(2004){Butler Burton}, {Braun}, \& {de
  Heij}}]{2004ASSL..312..313B}
{Butler Burton}, W., {Braun}, R., \& {de Heij}, V. 2004, {Compact, Isolated
  High-Velocity Clouds}, ed. H.~{van Woerden}, B.~P. {Wakker}, U.~J. {Schwarz},
  \& K.~S. {de Boer}, Vol. 312, 313

\bibitem[{{Dame} \& {Thaddeus}(2011)}]{2011ApJ...734L..24D}
{Dame}, T.~M., \& {Thaddeus}, P. 2011, \apjl, 734, L24

\bibitem[{{Digel} {et~al.}(1994){Digel}, {de Geus}, \&
  {Thaddeus}}]{1994ApJ...422...92D}
{Digel}, S., {de Geus}, E., \& {Thaddeus}, P. 1994, \apj, 422, 92

\bibitem[{{Dobbs} \& {Bonnell}(2006)}]{2006MNRAS.367..873D}
{Dobbs}, C.~L., \& {Bonnell}, I.~A. 2006, \mnras, 367, 873

\bibitem[{{Duarte-Cabral} \& {Dobbs}(2016)}]{2016MNRAS.458.3667D}
{Duarte-Cabral}, A., \& {Dobbs}, C.~L. 2016, \mnras, 458, 3667

\bibitem[{{Fiege} \& {Pudritz}(2000)}]{2000MNRAS.311...85F}
{Fiege}, J.~D., \& {Pudritz}, R.~E. 2000, \mnras, 311, 85

\bibitem[{{Fischera} \& {Martin}(2012)}]{2012A&A...542A..77F}
{Fischera}, J., \& {Martin}, P.~G. 2012, \aap, 542, A77

\bibitem[{{Gong} {et~al.}(2018){Gong}, {Li}, {Mao}, {Henkel}, {Menten}, {Fang},
  {Wang}, \& {Sun}}]{2018A&A...620A..62G}
{Gong}, Y., {Li}, G.~X., {Mao}, R.~Q., {et~al.} 2018, \aap, 620, A62

\bibitem[{{Goodman} {et~al.}(2014){Goodman}, {Alves}, {Beaumont}, {Benjamin},
  {Borkin}, {Burkert}, {Dame}, {Jackson}, {Kauffmann}, {Robitaille}, \&
  {Smith}}]{2014ApJ...797...53G}
{Goodman}, A.~A., {Alves}, J., {Beaumont}, C.~N., {et~al.} 2014, \apj, 797, 53

\bibitem[{{Green}(2014)}]{2014BASI...42...47G}
{Green}, D.~A. 2014, Bulletin of the Astronomical Society of India, 42, 47

\bibitem[{{Heitsch} \& {Putman}(2009)}]{2009ApJ...698.1485H}
{Heitsch}, F., \& {Putman}, M.~E. 2009, \apj, 698, 1485

\bibitem[{{HI4PI Collaboration} {et~al.}(2016){HI4PI Collaboration}, {Ben
  Bekhti}, {Fl{\"o}er}, {Keller}, {Kerp}, {Lenz}, {Winkel}, {Bailin},
  {Calabretta}, {Dedes}, {Ford}, {Gibson}, {Haud}, {Janowiecki}, {Kalberla},
  {Lockman}, {McClure-Griffiths}, {Murphy}, {Nakanishi}, {Pisano}, \&
  {Staveley-Smith}}]{2016A&A...594A.116H}
{HI4PI Collaboration}, {Ben Bekhti}, N., {Fl{\"o}er}, L., {et~al.} 2016, \aap,
  594, A116

\bibitem[{{Inutsuka} \& {Miyama}(1992)}]{1992ApJ...388..392I}
{Inutsuka}, S.-I., \& {Miyama}, S.~M. 1992, \apj, 388, 392

\bibitem[{{Inutsuka} \& {Miyama}(1997)}]{1997ApJ...480..681I}
{Inutsuka}, S.-i., \& {Miyama}, S.~M. 1997, \apj, 480, 681

\bibitem[{{Jackson} {et~al.}(2010){Jackson}, {Finn}, {Chambers}, {Rathborne},
  \& {Simon}}]{2010ApJ...719L.185J}
{Jackson}, J.~M., {Finn}, S.~C., {Chambers}, E.~T., {Rathborne}, J.~M., \&
  {Simon}, R. 2010, \apjl, 719, L185

\bibitem[{{Jiang} {et~al.}(2020){Jiang}, {Tang}, {Hou}, {Liu}, {Kr{\v{c}}o},
  {Qian}, {Sun}, {Ching}, {Liu}, {Duan}, {Yue}, {Gan}, {Yao}, {Li}, {Pan},
  {Yu}, {Liu}, {Li}, {Peng}, {Yan}, \& {FAST
  Collaboration}}]{2020RAA....20...64J}
{Jiang}, P., {Tang}, N.-Y., {Hou}, L.-G., {et~al.} 2020, Research in Astronomy
  and Astrophysics, 20, 064

\bibitem[{{Kalberla} {et~al.}(2007){Kalberla}, {Dedes}, {Kerp}, \&
  {Haud}}]{2007A&A...469..511K}
{Kalberla}, P.~M.~W., {Dedes}, L., {Kerp}, J., \& {Haud}, U. 2007, \aap, 469,
  511

\bibitem[{{Kalberla} \& {Haud}(2015)}]{2015A&A...578A..78K}
{Kalberla}, P.~M.~W., \& {Haud}, U. 2015, \aap, 578, A78

\bibitem[{{Kalberla} \& {Kerp}(2009)}]{2009ARA&A..47...27K}
{Kalberla}, P. M.~W., \& {Kerp}, J. 2009, \araa, 47, 27

\bibitem[{{Kalberla} {et~al.}(2020){Kalberla}, {Kerp}, \&
  {Haud}}]{2020A&A...639A..26K}
{Kalberla}, P.~M.~W., {Kerp}, J., \& {Haud}, U. 2020, \aap, 639, A26

\bibitem[{{Kalberla} {et~al.}(2016){Kalberla}, {Kerp}, {Haud}, {Winkel}, {Ben
  Bekhti}, {Fl{\"o}er}, \& {Lenz}}]{2016ApJ...821..117K}
{Kalberla}, P.~M.~W., {Kerp}, J., {Haud}, U., {et~al.} 2016, \apj, 821, 117

\bibitem[{{Kalberla} {et~al.}(2010){Kalberla}, {McClure-Griffiths}, {Pisano},
  {Calabretta}, {Ford}, {Lockman}, {Staveley-Smith}, {Kerp}, {Winkel},
  {Murphy}, \& {Newton-McGee}}]{2010A&A...521A..17K}
{Kalberla}, P.~M.~W., {McClure-Griffiths}, N.~M., {Pisano}, D.~J., {et~al.}
  2010, \aap, 521, A17

\bibitem[{{Kerp} {et~al.}(2011){Kerp}, {Winkel}, {Ben Bekhti}, {Fl{\"o}er}, \&
  {Kalberla}}]{2011AN....332..637K}
{Kerp}, J., {Winkel}, B., {Ben Bekhti}, N., {Fl{\"o}er}, L., \& {Kalberla},
  P.~M.~W. 2011, Astronomische Nachrichten, 332, 637

\bibitem[{{Kwak} {et~al.}(2011){Kwak}, {Henley}, \&
  {Shelton}}]{2011ApJ...739...30K}
{Kwak}, K., {Henley}, D.~B., \& {Shelton}, R.~L. 2011, \apj, 739, 30

\bibitem[{{Levine} {et~al.}(2006){Levine}, {Blitz}, \&
  {Heiles}}]{2006ApJ...643..881L}
{Levine}, E.~S., {Blitz}, L., \& {Heiles}, C. 2006, \apj, 643, 881

\bibitem[{{Li} {et~al.}(2018{\natexlab{a}}){Li}, {Wang}, {Zhang}, {Ma}, {Fang},
  \& {Yang}}]{2018ApJS..238...10L}
{Li}, C., {Wang}, H., {Zhang}, M., {et~al.} 2018{\natexlab{a}}, \apjs, 238, 10

\bibitem[{{Li} {et~al.}(2020){Li}, {Wang}, {Zhang}, {Ma}, \&
  {Lin}}]{2020ApJS..249...27L}
{Li}, C., {Wang}, H., {Zhang}, M., {Ma}, Y., \& {Lin}, L. 2020, \apjs, 249, 27

\bibitem[{{Li} {et~al.}(2018{\natexlab{b}}){Li}, {Wang}, {Qian}, {Krco},
  {Jiang}, {Yue}, {Jin}, {Zhu}, {Pan}, {Nan}, \&
  {Dunning}}]{2018IMMag..19..112L}
{Li}, D., {Wang}, P., {Qian}, L., {et~al.} 2018{\natexlab{b}}, IEEE Microwave
  Magazine, 19, 112

\bibitem[{{Lockman}(2003)}]{2003ApJ...591L..33L}
{Lockman}, F.~J. 2003, \apjl, 591, L33

\bibitem[{{Matsuo} {et~al.}(2017){Matsuo}, {Nakanishi}, {Minamidani}, {Torii},
  {Saito}, {Kuno}, {Sawada}, {Tosaki}, {Kobayashi}, {Yasui}, {Mito},
  {Hasegawa}, \& {Hirota}}]{2017PASJ...69L...3M}
{Matsuo}, M., {Nakanishi}, H., {Minamidani}, T., {et~al.} 2017, \pasj, 69, L3

\bibitem[{{McClure-Griffiths} {et~al.}(2009){McClure-Griffiths}, {Pisano},
  {Calabretta}, {Ford}, {Lockman}, {Staveley-Smith}, {Kalberla}, {Bailin},
  {Dedes}, {Janowiecki}, {Gibson}, {Murphy}, {Nakanishi}, \&
  {Newton-McGee}}]{2009ApJS..181..398M}
{McClure-Griffiths}, N.~M., {Pisano}, D.~J., {Calabretta}, M.~R., {et~al.}
  2009, \apjs, 181, 398

\bibitem[{{McConnachie}(2012)}]{2012AJ....144....4M}
{McConnachie}, A.~W. 2012, \aj, 144, 4

\bibitem[{{Men'shchikov} {et~al.}(2010){Men'shchikov}, {Andr{\'e}}, {Didelon},
  {K{\"o}nyves}, {Schneider}, {Motte}, {Bontemps}, {Arzoumanian}, {Attard},
  {Abergel}, {Baluteau}, {Bernard}, {Cambr{\'e}sy}, {Cox}, {di Francesco}, {di
  Giorgio}, {Griffin}, {Hargrave}, {Huang}, {Kirk}, {Li}, {Martin}, {Minier},
  {Miville-Desch{\^e}nes}, {Molinari}, {Olofsson}, {Pezzuto}, {Roussel},
  {Russeil}, {Saraceno}, {Sauvage}, {Sibthorpe}, {Spinoglio}, {Testi},
  {Ward-Thompson}, {White}, {Wilson}, {Woodcraft}, \&
  {Zavagno}}]{2010A&A...518L.103M}
{Men'shchikov}, A., {Andr{\'e}}, P., {Didelon}, P., {et~al.} 2010, \aap, 518,
  L103

\bibitem[{{Myers}(2009)}]{2009ApJ...700.1609M}
{Myers}, P.~C. 2009, \apj, 700, 1609

\bibitem[{{Panopoulou} {et~al.}(2017){Panopoulou}, {Psaradaki}, {Skalidis},
  {Tassis}, \& {Andrews}}]{2017MNRAS.466.2529P}
{Panopoulou}, G.~V., {Psaradaki}, I., {Skalidis}, R., {Tassis}, K., \&
  {Andrews}, J.~J. 2017, \mnras, 466, 2529

\bibitem[{{Putman} {et~al.}(2012){Putman}, {Peek}, \&
  {Joung}}]{2012ARA&A..50..491P}
{Putman}, M.~E., {Peek}, J.~E.~G., \& {Joung}, M.~R. 2012, \araa, 50, 491

\bibitem[{{Putman} {et~al.}(2011){Putman}, {Saul}, \&
  {Mets}}]{2011MNRAS.418.1575P}
{Putman}, M.~E., {Saul}, D.~R., \& {Mets}, E. 2011, \mnras, 418, 1575

\bibitem[{{Ragan} {et~al.}(2014){Ragan}, {Henning}, {Tackenberg}, {Beuther},
  {Johnston}, {Kainulainen}, \& {Linz}}]{2014A&A...568A..73R}
{Ragan}, S.~E., {Henning}, T., {Tackenberg}, J., {et~al.} 2014, \aap, 568, A73

\bibitem[{{Reid} {et~al.}(2016){Reid}, {Dame}, {Menten}, \&
  {Brunthaler}}]{2016ApJ...823...77R}
{Reid}, M.~J., {Dame}, T.~M., {Menten}, K.~M., \& {Brunthaler}, A. 2016, \apj,
  823, 77

\bibitem[{{Reid} {et~al.}(2014){Reid}, {Menten}, {Brunthaler}, {Zheng}, {Dame},
  {Xu}, {Wu}, {Zhang}, {Sanna}, {Sato}, {Hachisuka}, {Choi}, {Immer},
  {Moscadelli}, {Rygl}, \& {Bartkiewicz}}]{2014ApJ...783..130R}
{Reid}, M.~J., {Menten}, K.~M., {Brunthaler}, A., {et~al.} 2014, \apj, 783, 130

\bibitem[{{Rygl} {et~al.}(2012){Rygl}, {Brunthaler}, {Sanna}, {Menten}, {Reid},
  {van Langevelde}, {Honma}, {Torstensson}, \&
  {Fujisawa}}]{2012A&A...539A..79R}
{Rygl}, K.~L.~J., {Brunthaler}, A., {Sanna}, A., {et~al.} 2012, \aap, 539, A79

\bibitem[{{Schneider} {et~al.}(2010){Schneider}, {Csengeri}, {Bontemps},
  {Motte}, {Simon}, {Hennebelle}, {Federrath}, \&
  {Klessen}}]{2010A&A...520A..49S}
{Schneider}, N., {Csengeri}, T., {Bontemps}, S., {et~al.} 2010, \aap, 520, A49

\bibitem[{{Schneider} {et~al.}(2012){Schneider}, {Csengeri}, {Hennemann},
  {Motte}, {Didelon}, {Federrath}, {Bontemps}, {Di Francesco}, {Arzoumanian},
  {Minier}, {Andr{\'e}}, {Hill}, {Zavagno}, {Nguyen-Luong}, {Attard},
  {Bernard}, {Elia}, {Fallscheer}, {Griffin}, {Kirk}, {Klessen}, {K{\"o}nyves},
  {Martin}, {Men'shchikov}, {Palmeirim}, {Peretto}, {Pestalozzi}, {Russeil},
  {Sadavoy}, {Sousbie}, {Testi}, {Tremblin}, {Ward-Thompson}, \&
  {White}}]{2012A&A...540L..11S}
{Schneider}, N., {Csengeri}, T., {Hennemann}, M., {et~al.} 2012, \aap, 540, L11

\bibitem[{{Schneider} {et~al.}(2013){Schneider}, {Csengeri}, {Hennemann},
  {Motte}, {Didelon}, {Federrath}, {Bontemps}, {Di Francesco}, {Arzoumanian},
  {Minier}, {Andr{\'e}}, {Hill}, {Zavagno}, {Nguyen-Luong}, {Attard},
  {Bernard}, {Elia}, {Fallscheer}, {Griffin}, {Kirk}, {Klessen}, {K{\"o}nyves},
  {Martin}, {Men'shchikov}, {Palmeirim}, {Peretto}, {Pestalozzi}, {Russeil},
  {Sadavoy}, {Sousbie}, {Testi}, {Tremblin}, {Ward-Thompson}, \&
  {White}}]{2013A&A...551C...1S}
---. 2013, \aap, 551, C1

\bibitem[{{Smith} {et~al.}(2014{\natexlab{a}}){Smith}, {Glover}, {Clark},
  {Klessen}, \& {Springel}}]{2014MNRAS.441.1628S}
{Smith}, R.~J., {Glover}, S. C.~O., {Clark}, P.~C., {Klessen}, R.~S., \&
  {Springel}, V. 2014{\natexlab{a}}, \mnras, 441, 1628

\bibitem[{{Smith} {et~al.}(2014{\natexlab{b}}){Smith}, {Glover}, \&
  {Klessen}}]{2014MNRAS.445.2900S}
{Smith}, R.~J., {Glover}, S. C.~O., \& {Klessen}, R.~S. 2014{\natexlab{b}},
  \mnras, 445, 2900

\bibitem[{{Soler} {et~al.}(2020){Soler}, {Beuther}, {Syed}, {Wang}, {Anderson},
  {Glover}, {Hennebelle}, {Heyer}, {Henning}, {Izquierdo}, {Klessen}, {Linz},
  {McClure-Griffiths}, {Ott}, {Ragan}, {Rugel}, {Schneider}, {Smith},
  {Sormani}, {Stil}, {Tre{\ss}}, \& {Urquhart}}]{2020A&A...642A.163S}
{Soler}, J.~D., {Beuther}, H., {Syed}, J., {et~al.} 2020, \aap, 642, A163

\bibitem[{{Su} {et~al.}(2015){Su}, {Zhang}, {Shao}, \&
  {Yang}}]{2015ApJ...811..134S}
{Su}, Y., {Zhang}, S., {Shao}, X., \& {Yang}, J. 2015, \apj, 811, 134

\bibitem[{{Sun} {et~al.}(2015){Sun}, {Xu}, {Yang}, {Li}, {Du}, {Zhang}, \&
  {Zhou}}]{2015ApJ...798L..27S}
{Sun}, Y., {Xu}, Y., {Yang}, J., {et~al.} 2015, \apjl, 798, L27

\bibitem[{{Suri} {et~al.}(2019){Suri}, {S{\'a}nchez-Monge}, {Schilke},
  {Clarke}, {Smith}, {Ossenkopf-Okada}, {Klessen}, {Padoan}, {Goldsmith},
  {Arce}, {Bally}, {Carpenter}, {Ginsburg}, {Johnstone}, {Kauffmann}, {Kong},
  {Lis}, {Mairs}, {Pillai}, {Pineda}, \& {Duarte-Cabral}}]{2019A&A...623A.142S}
{Suri}, S., {S{\'a}nchez-Monge}, {\'A}., {Schilke}, P., {et~al.} 2019, \aap,
  623, A142

\bibitem[{{Tumlinson} {et~al.}(2017){Tumlinson}, {Peeples}, \&
  {Werk}}]{2017ARA&A..55..389T}
{Tumlinson}, J., {Peeples}, M.~S., \& {Werk}, J.~K. 2017, \araa, 55, 389

\bibitem[{{Wakker}(1991)}]{1991A&A...250..499W}
{Wakker}, B.~P. 1991, \aap, 250, 499

\bibitem[{{Wang} {et~al.}(2015){Wang}, {Testi}, {Ginsburg}, {Walmsley},
  {Molinari}, \& {Schisano}}]{2015MNRAS.450.4043W}
{Wang}, K., {Testi}, L., {Ginsburg}, A., {et~al.} 2015, \mnras, 450, 4043

\bibitem[{{Wang} {et~al.}(2020){Wang}, {Beuther}, {Rugel}, {Soler}, {Stil},
  {Ott}, {Bihr}, {McClure-Griffiths}, {Anderson}, {Klessen}, {Goldsmith},
  {Roy}, {Glover}, {Urquhart}, {Heyer}, {Linz}, {Smith}, {Bigiel}, {Dempsey},
  \& {Henning}}]{2020A&A...634A..83W}
{Wang}, Y., {Beuther}, H., {Rugel}, M.~R., {et~al.} 2020, \aap, 634, A83

\bibitem[{{Wilson} {et~al.}(2009){Wilson}, {Rohlfs}, \&
  {H{\"u}ttemeister}}]{2009tra..book.....W}
{Wilson}, T.~L., {Rohlfs}, K., \& {H{\"u}ttemeister}, S. 2009, {Tools of Radio
  Astronomy}

\bibitem[{{Winkel} {et~al.}(2016){Winkel}, {Kerp}, {Fl{\"o}er}, {Kalberla},
  {Ben Bekhti}, {Keller}, \& {Lenz}}]{2016A&A...585A..41W}
{Winkel}, B., {Kerp}, J., {Fl{\"o}er}, L., {et~al.} 2016, \aap, 585, A41

\bibitem[{{Xiong} {et~al.}(2017){Xiong}, {Chen}, {Yang}, {Fang}, {Zhang},
  {Zhang}, {Du}, \& {Long}}]{2017ApJ...838...49X}
{Xiong}, F., {Chen}, X., {Yang}, J., {et~al.} 2017, \apj, 838, 49

\bibitem[{{Xu} {et~al.}(2013){Xu}, {Li}, {Reid}, {Menten}, {Zheng},
  {Brunthaler}, {Moscadelli}, {Dame}, \& {Zhang}}]{2013ApJ...769...15X}
{Xu}, Y., {Li}, J.~J., {Reid}, M.~J., {et~al.} 2013, \apj, 769, 15

\bibitem[{{Zhang} {et~al.}(2019){Zhang}, {Kainulainen}, {Mattern}, {Fang}, \&
  {Henning}}]{2019A&A...622A..52Z}
{Zhang}, M., {Kainulainen}, J., {Mattern}, M., {Fang}, M., \& {Henning}, T.
  2019, \aap, 622, A52

\bibitem[{{Zucker} {et~al.}(2015){Zucker}, {Battersby}, \&
  {Goodman}}]{2015ApJ...815...23Z}
{Zucker}, C., {Battersby}, C., \& {Goodman}, A. 2015, \apj, 815, 23

\bibitem[{{Zucker} {et~al.}(2018){Zucker}, {Battersby}, \&
  {Goodman}}]{2018ApJ...864..153Z}
---. 2018, \apj, 864, 153

\end{thebibliography}


\end{document}